\documentclass[
  reprint,
  superscriptaddress,
  amsmath,amssymb,
  aps,prx
]{revtex4-2}

\usepackage{txfontsb}
\usepackage[table,xcdraw]{xcolor}
\usepackage{graphicx}
\usepackage{dcolumn}
\usepackage{bm}
\usepackage{float}
\usepackage{algpseudocode}
\usepackage{algorithm}
\usepackage{physics}
\usepackage{placeins}
\usepackage{booktabs}
\usepackage{mathtools}
\usepackage{mathrsfs}

\usepackage{amsthm}
\usepackage{subcaption}
\usepackage{acro}

\usepackage{xr-hyper}
\usepackage[
  colorlinks=true,
  linkcolor=blue,
  citecolor=blue,
  urlcolor=blue
]{hyperref}
\usepackage[all]{hypcap}
\usepackage[capitalise]{cleveref}

\newcommand{\Ket}[1]{|#1\rangle}
\newcommand{\Bra}[1]{\langle #1|}

\newtheorem{definition}{Definition}

\DeclareAcronym{qml}{
  short = QML,
  long = Quantum Machine Learning
}
\DeclareAcronym{qrc}{
  short = QRC,
  long = Quantum Reservoir Computing
}

\DeclareAcronym{mc}{
  short = MC,
  long = Memory Capacity
}
\DeclareAcronym{fmp}{
  short = FMP,
  long = Fading Memory Property
}
\DeclareAcronym{ipc}{
  short = IPC,
  long = Information Processing Capacity
}

\AtBeginDocument{%
    \setlength{\abovedisplayskip}{1pt plus 1pt minus 1pt}
    \setlength{\belowdisplayskip}{1pt plus 1pt minus 1pt}
    \setlength{\abovedisplayshortskip}{1pt plus 1pt minus 1pt}
    \setlength{\belowdisplayshortskip}{1pt plus 1pt minus 1pt}
}

\newcommand{\UTOKYO}{%
  \affiliation{Department of Applied Physics,
  the University of Tokyo, 7-3-1 Hongo,
  Bunkyo-ku, Tokyo 113-8656, Japan}%
}

\newcommand{\ILMENAU}{%
  \affiliation{Institute of Physics,
  Technische Universit\"at Ilmenau, Ilmenau, Germany}%
}

\newcommand{\ITPTUB}{%
  \affiliation{Institut f\"ur Physik und Astronomie,
  Technische Universit\"at Berlin,
  Hardenbergstr.\ 36, D-10623 Berlin, Germany}
}

\begin{document}

\title{Data re-uploading in Quantum Reservoir Computing}

\author{Saud \v{C}indrak}
\email{saud.cindrak@tu-ilmenau.de}
\ILMENAU

\author{Kaito Kobayashi}
\UTOKYO

\author{Lara Giebeler}
\ITPTUB

\author{Kathy L\"{u}dge}
\ILMENAU

\date{\today}

\begin{abstract}
Data re-uploading enhances the learnable function space of parameterized quantum circuits and has become standard practice in circuit-based \ac{qml}. Its role in quantum reservoir computing, however, remains largely unexplored.
Here, we introduce data re-uploading into physical quantum reservoir computing and show that repeated input encoding significantly enhances the learnability of higher-order nonlinear polynomials and improves performance on chaotic time-series prediction tasks.
We consider protocols with either fixed or increasing computational cost and compare them with a complementary dissipative protocol to distinguish the effects of repeated input encoding from those of additional state overwriting.
Using the information processing capacity framework, we identify a memory--nonlinearity trade-off: increasing the number of encodings shifts capacity toward higher-order nonlinearities.
In contrast, the dissipative protocol does not extend the range of learnable nonlinear orders and yields only limited improvements in task performance. These results establish data re-uploading as an effective means of tuning and enhancing nonlinear processing in physical quantum reservoir computing.
\end{abstract}

\maketitle

\textit{Introduction---}  \ac{qrc} has emerged as a subfield in \ac{qml}, which aims to harness the exponentially growing Hilbert space and complex dynamics of quantum systems for information processing \cite{FUJ17}. In contrast to variational gate-based quantum machine learning, where the parameters of a quantum circuit are optimized for a specific task \cite{CER22}, \ac{qrc} exploits the intrinsic dynamics of a quantum system and restricts training to a classical readout layer, which makes \ac{qrc} a promising approach for utilizing currently available quantum systems and thus has gained significant attention in recent years \cite{SUZ22, NAK19, GAR23, MUJ21a, ABB24a, FUJ21, NAK20, PFE23, KAW26, KOB24a, GOE23, MUJ23, DOM24, CIN24, CIN24, CIN25a, CIN25, Hou2026, Kornjaca2024, Paparelle2026, Chen2020}.
Within \ac{qml}, parameterized quantum circuits provide an approach based on trainable quantum circuits whose parameters are optimized for a given task~\cite{CER22, ARU19, SCH15a, BIA17, HAV19, SCH21a}. The functions accessible to such models are mainly influenced by the chosen data-encoding strategy~\cite{SCH19}. To increase their expressivity, Pérez-Salinas et al.~\cite{PER20} introduced data re-uploading, in which the input is repeatedly encoded throughout the quantum circuit by interleaving data-encoding operations with parameterized quantum operations. This repeated encoding has been shown to significantly increase the number of learnable functions~\cite{PER20,SCH21} and has thus become standard practice in circuit-based \ac{qml}~\cite{PER20,SCH21,PER21,SCH22,JER23,ONO23,CON26,WAC23,BAR24,COE24}, but remains largely unexplored in the context of physical \ac{qrc}.

In particular, it remains unclear to what extent repeated input encoding affects the task performance and information-processing capabilities of quantum reservoir computers, and whether the success of data re-uploading in \ac{qml} can be translated to physical \ac{qrc}.
Where in conventional circuit-based data re-uploading, additional encoding layers increase both the number of encoding operations and the circuit depth, extending the total evolution time. In physical \ac{qrc}, the quantum system instead evolves over a fixed clock cycle $T$. Data can be re-uploaded at intermediate times within this interval, allowing $T$ to remain fixed regardless of the number of re-uploads. 
The resulting increase in execution time is therefore due solely to additional encoding operations and is thus a distinct advantage of physical \ac{qrc}.  
%The ability to increase the number of re-uploads without extending the evolution time is a distinct advantage of physical \ac{qrc}. For completeness, we also consider a protocol in which the interval between successive encodings is fixed, more closely resembling conventional data re-uploading in circuit-based \ac{qml} \cite{PER20, XIO23}. 
To ensure that any observed improvement in task performance is due to data re-uploading rather than a partial state reset, we compare the data re-uploading protocol with a dissipative protocol in which the encoding qubit is reset to a fixed state instead of re-encoding the input.  
We characterize these protocols using nonlinear time-series prediction tasks, namely the Lorenz system \cite{LOR63} and the Mackey Glass system \cite{MAC77}. To probe the learnable class of functions, we consider the \ac{ipc} framework \cite{DAM12}, which provides a task-independent decomposition of the reservoir's information-processing capabilities into contributions of different polynomial orders and has shed insights in classical \cite{BAU24, HUE22a, INU17, RAT25, HUA26, AUS25} and quantum reservoir computing \cite{MAR20, NOK21, NOK26, CIN24, CIN25, CIN25a, CIN26a}. 

Our results show that data re-uploading results in significant enhancements of higher-order \ac{ipc} contributions and improves task performance, even when the total evolution time is held fixed. At sufficiently high encoding frequencies, however, the computational performance deteriorates as the encoding qubit becomes increasingly frozen. 
In contrast, the dissipative protocol enhances existing nonlinearities through qubit resets but fails to capture higher-order nonlinearities. For all tasks considered, data re-uploading outperforms the dissipative protocol, identifying data re-uploading as a previously unexploited resource for physical \ac{qrc}.
To investigate whether data re-uploading improves performance under dissipative dynamics, we propose a hybrid protocol that interleaves input encoding and dissipation. Our results show that even limited data re-uploading improves performance under dissipation and that, under certain conditions, the hybrid protocol outperforms both individual protocols.

Taken together, these results demonstrate that data re-uploading offers an efficient and underexplored approach to enhancing and controlling system nonlinearities, highlighting its potential as a standard technique in physical \ac{qrc}, as already established in circuit-based \ac{qml}~\cite{PER20}.
%Together, these results establish data re-uploading as a broadly applicable yet underexplored tool for enhancing physical \ac{qrc}. 
 
\begin{figure}[t]
    \centering
    \includegraphics[
        angle=0,
        scale=1
    ]{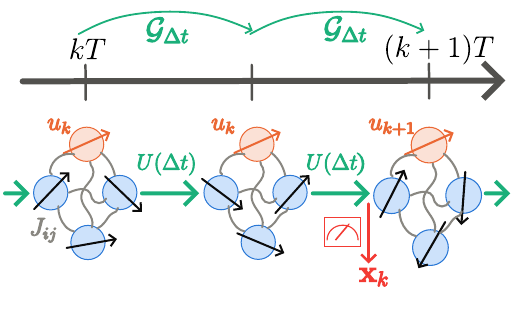}
    \caption{Data re-uploading scheme for physical \ac{qrc} with $n_{\mathrm{enc}}=2$. The operator $U(\Delta t)$ describes unitary evolution over a time interval $\Delta t$, while $\mathcal{G}_{\Delta t}$ combines one encoding step with the subsequent unitary evolution. The input $s_k$ is encoded twice per clock cycle, at $kT$ and $(k+\tfrac{1}{2})T$.}
    \label{fig:scheme}
\end{figure} 

\textit{Quantum Reservoir Computing---} We briefly review how data re-uploading is employed in \ac{qml}. In a typical variational \ac{qml} model, the output state is given by
$
\ket{y(\theta,x)} = U(\theta)U_E(x)\ket{0},
$
where \(U(\theta)\) denotes a parameterized quantum circuit with trainable parameters \(\theta\in\mathbb{R}^{n_P}\), \(U_E(x)\) is the data-encoding unitary for the classical input \(x\), and \(\ket{0}\) denotes the initial quantum state.
The model output is obtained as the expectation value of an observable \(O\),
$
f_\theta(x)=\bra{y(\theta,x)}O\ket{y(\theta,x)}.
$
The corresponding accessible function class 
\(\mathcal{F}_{U_E}=\{f_\theta:x\mapsto f_\theta(x)\mid\theta\in\mathbb{R}^{n_P}\}\)
and the nonlinearity in the input data only depends on the choice of the data-encoding strategy \(U_E(x)\)~\cite{SCH19}. 
To increase the accessible function space $\mathcal{F}_{U_E}$ of the model, data re-uploading repeatedly interleaves the encoding of the same input with parameterized quantum operations, resulting in the output state~
$
\ket{y(\theta,x)}
=
\left(
\prod_{k=1}^{n_{\mathrm{enc}}}
U(\theta_k)U_E(x)
\right)\ket{0} ~\cite{PER20, SCH21}.
$

To employ data re-uploading within \ac{qrc}, we first revisit physical \ac{qrc} as introduced by Fujii et al.~\cite{FUJ17}. In physical \ac{qrc}, we consider a discrete input time series \((u_k)_{k\in\mathbb{Z}}\).
At the $k$th time step, the \ac{qrc} performs the input-dependent map $\rho_k=
\mathcal{G}_{T}(\rho_{k-1})$, with 
\begin{align}
\mathcal{G}_{T}(\rho_{k-1})
=
e^{-iHT}
\left(
\Ket{\psi_k^{(1)}}\Bra{\psi_k^{(1)}}
\otimes
\Tr_1[\rho_{k-1}]
\right)
e^{iHT},
\label{eq:map_G}
\end{align}
where $\rho_k$ denotes the state of the full system,
\(
\Ket{\psi_k^{(1)}}
=
U_E(u_k)\Ket{0}
=
\sqrt{\frac{1-u_k}{2}}\Ket{0}
+
\sqrt{\frac{1+u_k}{2}}\Ket{1}
\)
denotes the encoded input state, $\Tr_1$ denotes the partial trace over the first qubit, $H$ is the reservoir Hamiltonian, and $T$ is the reservoir clock cycle \cite{FUJ17}.

\begin{figure}[t]
    \centering
    \includegraphics[scale = 1]{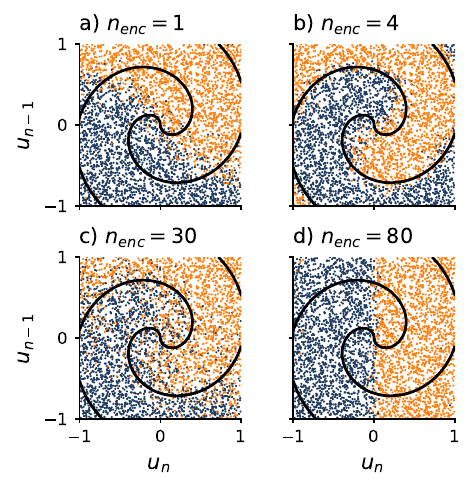}
    \caption{Spiral classification task for  $n_{\mathrm{enc}} =1$ in (a), $n_{\mathrm{enc}} =4$ in (b), $n_{\mathrm{enc}} =30$ in (c) and $n_{\mathrm{enc}} =80$ in (d).  }
    \label{fig:0_spiral}
\end{figure}

Whereas data re-uploading in \ac{qml} increases the circuit depth and, consequently, the execution time of the algorithm, physical \ac{qrc} relies on the natural evolution of the quantum system over a time interval \(T\), which serves as a measure of computational cost. To introduce data re-uploading without increasing this cost, we keep \(T=\mathrm{const.}\) and divide each clock cycle into \(n_{\mathrm{enc}}\) intervals of duration
\(
\Delta t = {T}/{n_{\mathrm{enc}}}.
\)
The same input is then repeatedly encoded at \(t_n=nT/n_{\mathrm{enc}}\), for \(n=0,\ldots,n_{\mathrm{enc}}-1\), with the reservoir evolving for \(\Delta t\) between successive encodings. The reservoir state after one clock cycle is then
\begin{align}
\hspace{-0.15cm}
\rho_k^{(n_{\mathrm{enc}})}
&=
\mathcal{G}_{\Delta t}^{\,n_{\mathrm{enc}}}(\rho_{k-1})
=
\underbrace{
\mathcal{G}_{\Delta t}
\circ
\mathcal{G}_{\Delta t}
\circ\cdots\circ
\mathcal{G}_{\Delta t}
}_{n_{\mathrm{enc}}\ \mathrm{times}}
(\rho_{k-1}).
\label{eq:map_G_collision}
\end{align}
An illustrative sketch of this protocol for the $k$th input $u_k$ with $n_{\mathrm{enc}}=2$ is shown in \cref{fig:scheme}. The input $u_k$ is encoded at $kT$ and $(k+\tfrac{1}{2})T$, with the system evolving between encoding steps. At the end of the clock cycle, $(k+1)T$, the system is measured to construct the readout vector $\mathbf{x}_k$. 

\begin{figure*}[t]
    \centering
    \includegraphics[scale=1]{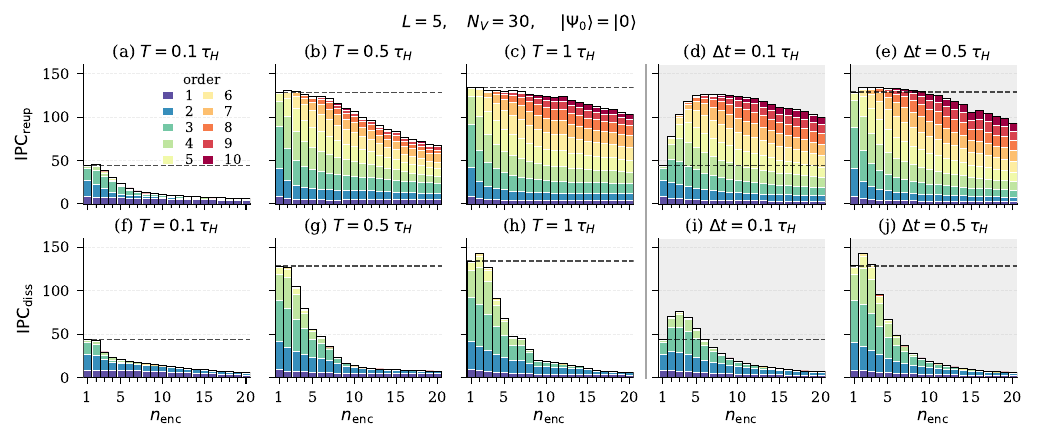}
    \caption{Information processing capacity (IPC) as a function of $n_{\mathrm{enc}}$ for the data re-uploading protocol (first row) and the dissipative protocol (second row) for fixed clock cycles (a--c) and (f--h) and fixed encoding intervals (d,e, i,j).
    \vspace{-7mm}}
    \label{fig:IPC_task}
\end{figure*}
We also consider a second protocol with fixed \(\Delta t\) and increasing clock cycle \(T=n_{\mathrm{enc}}\Delta t\), more closely resembling conventional data re-uploading in \ac{qml}~\cite{PER20}.
To isolate the effect of re-encoding, we also consider a protocol in which we reinitialize the state as $\Ket{\psi_k^{(1)}}=\Ket{\psi_0}=\Ket{0}$. This introduces additional dissipation without providing any additional information about the input, allowing us to identify improvements in task performance that arise specifically from data re-uploading.
In summary, we consider four protocols, combining either data re-uploading or dissipation with either a fixed clock cycle, $T=\mathrm{const.}$, or an increasing clock cycle, $T=n_{\mathrm{enc}}\Delta t$.
We consider a transverse-field Ising Hamiltonian with disordered interactions,
\begin{align}
H
=
\frac{1}{2}
\sum_{i=1}^{L}\sum_{j>i}^{L}
J_{ij}\sigma_x^{(i)}\sigma_x^{(j)}
+
\frac{1}{2}
\sum_{i=1}^{L}\sigma_z^{(i)},
\label{eq:hamiltonian}
\end{align}
as a quantum reservoir, where \(\sigma_x^{(i)}\) and \(\sigma_z^{(i)}\) denote the Pauli-\(x\) and Pauli-\(z\) operators acting on the \(i\)th site, respectively. We average all results over ten independent realizations of the Hamiltonian, with each coupling \(J_{ij}\) independently sampled from the uniform distribution \(\mathcal{U}([0.25,0.75])\).

Finally, a \ac{qrc} collects its readout by measuring \(K\) observables \((O_i)_{i=1}^{K}\), in our case $O_i=\sigma_z^{(i)}$, at \(N_V\) intermediate times \(t_n=nT/N_V\) within each clock cycle. This results in a readout vector \(\mathbf{x}_k\) of dimension \(N_R=KN_V\) for each input \(u_k\). A classical linear readout is then trained such that \(\hat{y}_k=\mathbf{x}_k W_{\mathrm{opt}}\) best approximates the target value \(y_k\). Further details are provided in App.~\ref{app:qrc}.

%%%%%%%%%%%%%%%%%%%%%%%%%%%%%%%%%%%%%%%%%%%%%%%%%%%%%%%%%%%
%%%%%%%%%%%%%%%%%%%%%%%%%%%%%%%%%%%%%%%%%%%%%%%%%%%%%%%%%%%
%%%%%%%%%%%%%%%%%%%%%%%%%%%%%%%%%%%%%%%%%%%%%%%%%%%%%%%%%%%

To illustrate the effect of data re-uploading at a fixed algorithmic execution time, we first consider a nonlinear classification task using an $L=4$ qubit reservoir (\cref{fig:0_spiral}). A time series of inputs $u_n$ drawn uniformly from $[-1,1]$ is sequentially encoded into the reservoir. The task is to assign each input pair $(u_n,u_{n-1})$ to one of the two spiral regions shown in \cref{fig:0_spiral}. This decision boundary therefore tests both the reservoir's ability to retain the previous input $u_{n-1}$ and its capacity to perform nonlinear transformations of the current and previous inputs.
The vanilla \ac{qrc} approach uses a single input encoding, $n_{\mathrm{enc}}=1$, and achieves a classification accuracy of $71\%$. Increasing the number of encodings to $n_{\mathrm{enc}}=4$ improves the separation and raises the accuracy to $88\%$. Further increases, however, reverse this trend: performance deteriorates at $n_{\mathrm{enc}}=30$, and at $n_{\mathrm{enc}}=80$, the reservoir effectively behaves as a linear classifier. This suggests an interplay between data re-uploading and frequent state overwrites acting as dissipation, motivating further investigation.

To provide such a systematic characterization, we employ established dynamical systems theory and reservoir computing, namely the \ac{mc}, \ac{fmp}, and \ac{ipc} \cite{DAM12,JAE01}. The \ac{mc} quantifies the ability of the reservoir to retain information about past inputs, the \ac{fmp} characterizes the fading of information associated with different initial conditions, and the \ac{ipc} measures the reservoir's capacity to reconstruct linear and nonlinear functions of previous inputs. Their definitions are provided in App.~\ref{app:ipc}.

%%%%%%%%%%%%%%%%%%%%%%%%%%%%%%%%%%%%%%%%%%%%%%%%%%%%%%%%
%%%%%%%%%%%%%% Individual Contributions %%%%%%%%%%%%%%%%
%%%%%%%%%%%%%%%%%%%%%%%%%%%%%%%%%%%%%%%%%%%%%%%%%%%%%%%% 

\textit{Results---} \Cref{fig:IPC_task} shows the \ac{ipc}s up to tenth order for the data re-uploading (first row) and dissipative (second row) protocols in an $L=5$ qubit system. For the dissipative protocol, we consider resets to the state $\Ket{\psi_k^{(1)}}=\Ket{0}$. Results for other reset states are provided in Supplemental Material \cite{Supp}, Sec. II.
In \cref{fig:IPC_task}(a--c), we consider the \ac{qrc} protocol with a fixed algorithmic execution time by keeping the clock cycle fixed at \(T\in\{0.1\tau_H,0.5\tau_H,1.0\tau_H\}\). In \cref{fig:IPC_task}(d,e) (grey background), we instead fix \(\Delta t\in\{0.1\tau_H,0.5\tau_H\}\), which results in an increase in the overall execution time to \(T=n_{\mathrm{enc}}\Delta t\) and corresponds more closely to conventional data re-uploading in circuit-based \ac{qml}.

For the shortest clock cycle, $T=0.1\tau_H$, vanilla QRC exhibits contributions up to approximately fourth-order IPC, whereas data re-uploading [\cref{fig:IPC_task}(a)] results in a slight increase up to fifth order. At $n_{\mathrm{enc}}=2$, we observe an enhancement relative to vanilla QRC, while for larger $n_{\mathrm{enc}}$ the IPC rapidly decreases as $\Delta t$ becomes too short for substantial reservoir evolution. For the dissipative protocol [\cref{fig:IPC_task}(f)] a similar behavior is observed, however the lower-order IPCs remain more pronounced with additional encodings.
For the longer clock cycle $T=0.5\tau_H$ [\cref{fig:IPC_task}(b)], data re-uploading leads to a pronounced redistribution of IPC from lower- to higher-order contributions. While vanilla \ac{qrc} exhibits significant nonlinear capacities only up to fifth order, repeated encoding generates contributions extending up to eighth order. In contrast, the dissipative protocol [\cref{fig:IPC_task}(g)] does not generate higher-order nonlinearities and remains limited to fifth order, as in vanilla \ac{qrc}. At higher encoding counts, the expressivity of the dissipative protocol decreases significantly faster, as excessive dissipation degrades performance, consistent with previous observations on dissipation in \ac{qrc}~\cite{GOE25,CIN26}.
Further increasing the clock cycle to $T=\tau_H$ shifts the IPC toward even higher orders, with contributions extending up to tenth order for the data re-uploading protocol [\cref{fig:IPC_task}(c)]. Since we evaluate the IPC only up to tenth order, part of the apparent decrease in total IPC at large $n_{\mathrm{enc}}$ may result from capacity being transferred to even higher orders that are not included here.
Apart from a small increase in total \ac{ipc} above the vanilla \ac{qrc} value [\cref{fig:IPC_task}(h)], the behavior of the dissipative protocol remains largely unchanged compared with the case $T=0.5\tau_H$ [\cref{fig:IPC_task}(g)].

\begin{figure}[t] 
\hspace{-0.3cm}
    \includegraphics[scale=1]{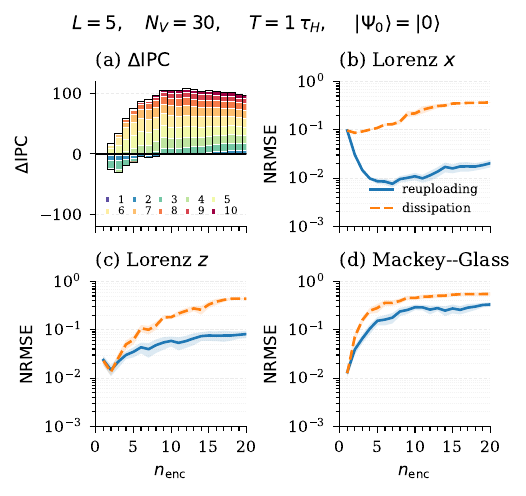} 
\caption{Difference in IPC between the data re-uploading and dissipative protocols (a), and task performance for Lorenz $x$ (b), Lorenz $z$ (c), and Mackey--Glass prediction (d).} 
    \label{fig:task}
\end{figure}
Next, we analyze the data re-uploading and dissipative protocols for fixed encoding intervals $\Delta t=0.1\tau_H$ [\cref{fig:IPC_task}(d,i)] and $\Delta t=0.5\tau_H$ [\cref{fig:IPC_task}(e,j)]. For $\Delta t=0.1\tau_H$, vanilla \ac{qrc} exhibits low expressivity due to the short encoding interval, consistent with the results shown in panel (a). Increasing the number of encodings, and thus the total evolution time $T$, significantly enhances the system's nonlinearities. In contrast, the dissipative protocol [\cref{fig:IPC_task}(i)] shows only a slight improvement, as the benefits of longer evolution times are partly offset by information loss due to repeated state overwrites, resulting in substantially lower expressivity than the data re-uploading protocol.
For data re-uploading with $\Delta t=0.5\tau_H$ [\cref{fig:IPC_task}(e)], we observe a redistribution toward higher-order capacities, similar to the case of $T=\tau_H$ [\cref{fig:IPC_task}(c)], revealing a memory--nonlinearity trade-off consistent with recent observations \cite{CIN26, KOB26} (see Supplemental Material \cite{Supp}, Sec. I for further details).
The dissipative protocol with $\Delta t=0.5\tau_H$ [\cref{fig:IPC_task}(j)] exhibits nearly identical qualitative behavior to the fixed-clock-cycle case $T=\tau_H$ [\cref{fig:IPC_task}(h)], indicating that clock cycles beyond the Heisenberg time provide little additional benefit \cite{CIN25, CIN26, KOB26}.
\begin{figure}[t]
    \centering
    \includegraphics[scale=1]{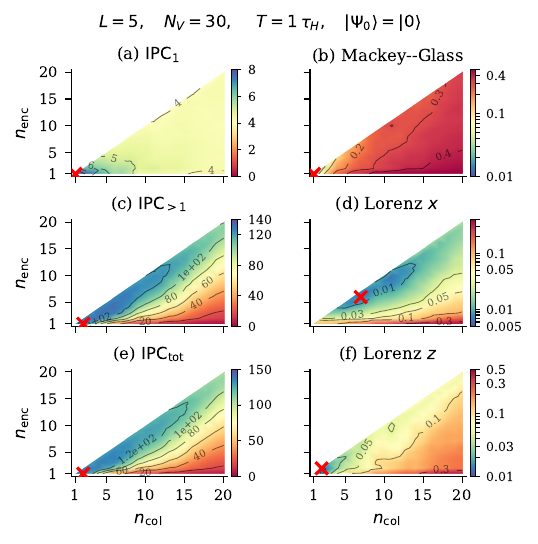}
    \caption{
    Performance landscapes of the hybrid protocol for
    $L=5$, $N_v=30$, and $T=\tau_H$.
    The panels show (a) linear IPC, (b) Mackey--Glass NRMSE,
    (c) nonlinear IPC, (d) Lorenz $x$ NRMSE,
    (e) total IPC, and (f) Lorenz $z$ NRMSE
    as functions of the total number of collisions $n_{\mathrm{col}}$
    and encodings $n_{\mathrm{enc}}$.
    Red crosses indicate the optimal configurations.
    }
    \label{fig:5_hybrid}
\end{figure}

%%%%%%%%%%%%%%%%%%%%%%%%%%%%%%%%%%%%%%%%%%%%%%%%%%%%%%%%
%%%%%%%%%%%%%%%%%% Task Performance  %%%%%%%%%%%%%%%%%%%
%%%%%%%%%%%%%%%%%%%%%%%%%%%%%%%%%%%%%%%%%%%%%%%%%%%%%%%% 
For the task evaluation, we first consider two prediction tasks based on the Lorenz system: predicting $x(t_{n+1})$ from $x(t_n)$ (Lorenz $x$, \cref{fig:task}(b)) and cross-predicting $z(t_n)$ from $x(t_n)$ (\cref{fig:task}(c)). We further evaluate the Mackey--Glass task (\cref{fig:task}(d)). The task definitions are provided in App.~\ref{app:tasks}. Prediction performance is quantified by the normalized root mean square error (NRMSE),
\begin{align}
    \mathrm{NRMSE} = \sqrt{\frac{\langle (y_n-\hat{y}_n)^2\rangle}{\mathrm{Var}(y_n)}},
\end{align}
with target values $y_n$ and predictions $\hat{y}_n$. We also compare the information processing capacities of the data re-uploading and dissipative protocols through their difference, $\Delta \mathrm{IPC}=\mathrm{IPC}_{\mathrm{reup}}-\mathrm{IPC}_{\mathrm{diss}}$ (\cref{fig:task}(a)).
These tasks place different demands on the reservoir's nonlinear processing and memory capabilities. The Lorenz $x$ task requires short memory and strong nonlinear processing, whereas the Lorenz $z$ task requires both longer memory and nonlinear processing due to the slower dynamics of the $z$-variable. The Mackey--Glass task is designed to impose the greatest memory requirements, exceeding the memory available in our reservoir and thereby limiting prediction performance. \\
\cref{fig:task}(a) shows that, in the initial regime, the dissipative protocol provides higher values for lower-order IPCs. However, as the number of encodings increases, this benefit is lost because the system increasingly forgets faster due to additional state override.
For the Lorenz $x$ task, data re-uploading leads to a significant improvement in task performance, whereas the dissipative protocol exhibits only a slight initial improvement followed by a pronounced deterioration [\cref{fig:task}(b)]. This behavior is consistent with the stronger higher-order IPC contributions generated by data re-uploading, indicating that the task benefits from enhanced nonlinear processing.
The Lorenz $z$ cross-prediction task involves slower dynamics and places a stronger emphasis on memory. Here, both the dissipative and data re-uploading protocols reach their minimum error at $n_{\mathrm{enc}}=2$, with nearly identical performance as shown in \cref{fig:task}(c). This suggests that, in this regime, the improvement is largely associated with the additional enhancement of lower-order nonlinearities in both protocols, as shown in \cref{fig:task}(a). 
Finally, the Mackey--Glass task represents a strongly memory-dependent setting. Here, any redistribution of capacity from the linear, memory-related contribution toward higher IPC orders leads to a deterioration in task performance [\cref{fig:task}(d)]. Consequently, vanilla QRC achieves the best performance among the protocols considered, demonstrating that data re-uploading is not universally beneficial when preserving memory is the dominant requirement.
These results demonstrate that data re-uploading substantially enhances higher-order nonlinearities and improves performance on nonlinear tasks, at the expense of reduced performance on memory-intensive tasks.

%%%%%%%%%%%%%%%%%%%%%%%%%%%%%%%%%%%%%%%%%%%%%%%%%%%%%%%%
%%%%%%%%%%%%%%%%% Hybrid Properties %%%%%%%%%%%%%%%%%%%%
%%%%%%%%%%%%%%%%%%%%%%%%%%%%%%%%%%%%%%%%%%%%%%%%%%%%%%%%
To investigate the interplay between data re-uploading and dissipation, we introduce a hybrid protocol that interleaves input encoding and state resets, continuously connecting the two limiting cases. We assess whether data re-uploading improves performance under dissipative dynamics and identify regimes in which the hybrid protocol outperforms both individual protocols. This further extends the analysis of data re-uploading to \ac{qrc} systems with different underlying dynamics.\\
For notation we consider $n_{\mathrm{col}}=n_{\mathrm{enc}}+n_{\mathrm{reset}}$ state overwrites or collisions, where $n_{\mathrm{reset}}$ denotes the number of qubit resets following the dissipative protocol, while $n_{\mathrm{enc}}$ denotes the number of input encodings. Collisions occur at equally spaced times $t_n=nT/n_{\mathrm{col}}$, with the initial encoding always performed at $t_0=0$. The additional data re-uploadings are distributed approximately uniformly over the remaining collision times, while all other collisions perform qubit resets. 
Figure~\ref{fig:5_hybrid} reveals a task-dependent optimal balance between data re-uploading and dissipation. Linear IPC and Mackey--Glass prediction perform best without additional collisions, whereas introducing a single reset maximizes nonlinear and total IPC. In contrast, Lorenz $x$ prediction favors repeated data re-uploading combined with only one reset, achieving its lowest prediction error at $n_{\mathrm{col}}=7$ and $n_{\mathrm{enc}}=6$, while Lorenz $z$ is optimal for $n_{\mathrm{enc}}=2 $ and without any dissipative state reset. 
The contour plots further reveal extended regions of high task performance for the hybrid, indicating that the optimal balance between data re-uploading and dissipation is not restricted to individual parameter combinations.
In \cref{sec:hybrid}, \cref{fig:app_hybrid} shows the individual IPC contributions and task performance for fixed $n_{\mathrm{col}}$. For $n_{\mathrm{col}}\in\{10,15\}$, the hybrid protocol outperforms both individual protocols on certain tasks, demonstrating that combining data re-uploading and dissipation can improve computational performance beyond either protocol alone. Additional results are given in Supplemental Material \cite{Supp}, Sec. III. 

We also consider a probabilistic hybrid protocol in which the input qubit is prepared in the mixed state $\rho=p\rho_{\mathrm{enc}}+(1-p)\rho_0$, where $\rho_0$ and $\rho_{\mathrm{enc}}$ denote the reset and encoded input states, respectively. Although this protocol can locally enhance the total IPC, pure data re-uploading consistently achieves the best task performance (see Supplemental Material \cite{Supp}, Sec. IV for the corresponding quantum circuit and additional results).

%%%%%%%%%%%%%%%%%%%%%%%%%%%%%%%%%%%%%%%%%%%%
%%%%%%%%%%%%%%%%%%%%%%%%%%%%%%%%%%%%%%%%%%%%

\textit{Conclusion---}
This work introduces data re-uploading into physical quantum reservoir computing and demonstrates that repeated input encoding provides a easy to adapt and flexible means of controlling the computational properties of the reservoir. In particular, data re-uploading substantially enhances higher-order information-processing capacities, extending the accessible nonlinearities beyond those of conventional QRC and the corresponding dissipative protocol. The information-processing-capacity framework further provides a quantitative characterization of the reservoir's learnable function space, offering a transparent way to relate the reservoir dynamics to its computational capabilities.
The enhanced nonlinear expressivity leads to significant improvements in task performance when strong nonlinear processing is required, whereas memory-dominated tasks can deteriorate as computational capacity is redistributed away from linear memory. 

To further investigate the interplay between data re-uploading and dissipation, we introduce a hybrid protocol that interleaves input encoding and state resets. This allows us to study data re-uploading under stronger dissipation while introducing an additional parameter for tuning the reservoir dynamics. Our results demonstrate that data re-uploading improves task performance under stronger dissipation and identify regions in which the hybrid protocol outperforms both individual protocols.

Overall, our results establish data re-uploading as an effective strategy for tuning the memory--nonlinearity trade-off in physical quantum reservoirs. Importantly, the resulting gains in nonlinear processing and task performance can already be achieved at fixed computational cost, making data re-uploading a simple and broadly applicable tool for enhancing existing QRC implementations.

\begin{acknowledgments} 
    S.\v{C}. acknowledges funding from the Deutsche Forschungsgemeinschaft (DFG, German Research Foundation) -- Project No. 588509319.\\
    K.K. was supported by the Program for Leading Graduate Schools (MERIT-WINGS) and JST BOOST (No.~JPMJBS2418). \\
    L.G. acknowledges funding by the Deutsche Forschungsgemeinschaft (DFG, German Research Foundation) via the Research Unit FOR 5688 (Project No. 521530974).
\end{acknowledgments}

\clearpage
\appendix
  
\section{Quantum Reservoir Computing}\label{app:qrc}

At each time step $k$, the input $u_k\in[-1,1]$ is encoded into the first qubit by replacing its state with
\begin{align}
\Ket{\psi_k^{(1)}}
=
\sqrt{\frac{1-u_k}{2}}\Ket{0}
+
\sqrt{\frac{1+u_k}{2}}\Ket{1}.
\end{align}
The remaining qubits retain information about previous inputs. Encoding followed by evolution under the reservoir Hamiltonian $H$ for a time interval $t$ defines the map
\begin{align}
\mathcal{G}_{t}(\rho)
=
e^{-iHt}
\left(
\Ket{\psi_k^{(1)}}\Bra{\psi_k^{(1)}}
\otimes
\Tr_1[\rho]
\right)
e^{iHt},
\end{align}
where $\Tr_1$ denotes the partial trace over the first qubit and we set $\hbar=1$.

In vanilla \ac{qrc}, the input is encoded once per clock cycle, yielding $\rho_k=\mathcal{G}_{T}(\rho_{k-1})$.
With data re-uploading, the same input $u_k$ is encoded $n_{\mathrm{enc}}$ times, with evolution intervals $\Delta t$ between successive encodings. The state at the end of the clock cycle is therefore
\begin{align}
\rho_k
=
\mathcal{G}_{\Delta t}^{\,n_{\mathrm{enc}}}
(\rho_{k-1}),
\qquad
T=n_{\mathrm{enc}}\Delta t,
\end{align}
where the power denotes repeated composition with the same input.
We consider either a fixed clock cycle $T$, for which $\Delta t=T/n_{\mathrm{enc}}$, or a fixed interval $\Delta t$, for which $T$ increases with $n_{\mathrm{enc}}$.
Information is extracted from the reservoir through the expectation values of \(K\) observables,
\begin{align}
\langle O_i(t)\rangle
=
\mathrm{Tr}\!\left[\rho_k(t)O_i\right],
\qquad i=1,\ldots,K.
\end{align}
To account for shot noise, we add Gaussian noise $\eta$ with amplitude $\alpha = 10^{-6}$ to each observable, $\langle O_k(T)\rangle \leftarrow \langle O_k(T)\rangle + \alpha\eta$, where $\eta \sim \mathcal{N}(0,1)$.
The expectation values at the end of each clock cycle define the readout vector
\begin{align}
\mathbf{x}_k
=
\left(
\langle O_1(T)\rangle,\ldots,\langle O_K(T)\rangle
\right).
\end{align}
The readout dimension can be increased either by measuring additional observables or through temporal multiplexing. In the latter case, the observables are sampled at intermediate times
\begin{align}
t_n=\frac{nT}{N_V},
\qquad n=1,\ldots,N_V,
\end{align}
such that each observable contributes \(N_V\) virtual nodes and the resulting readout dimension is
\begin{align}
N_R=KN_V.
\end{align}
In this work, we consider \(O_i=\sigma_z^{(i)}\) as the measured observables and use \(N_V=30\) temporal multiplexing points. For the four-spin reservoir, this results in \(K=4\) observables and a total readout dimension of \(N_R=120\)~\cite{FUJ17}.

For an input sequence of length \(N_T\), the readout vectors are collected row-wise in the feature matrix
\begin{align}
\mathbf{X}
=
\begin{pmatrix}
\mathbf{x}_1\\
\mathbf{x}_2\\
\vdots\\
\mathbf{x}_{N_T}
\end{pmatrix}
\in\mathbb{R}^{N_T\times N_R},
\end{align}
where the \(k\)th row corresponds to the reservoir response \(\mathbf{x}_k\). Given the target vector
\begin{align}
\mathbf{y}
=
(y_1,\ldots,y_{N_T})^{\mathrm{T}},
\end{align}
the reservoir prediction is obtained through the linear readout
\begin{align}
\hat{\mathbf{y}}
=
\mathbf{X}W,
\end{align}
where \(W\) contains the trainable readout weights. The readout is trained using Tikhonov-regularized least squares, yielding
\begin{align}
W_{\mathrm{opt}}
=
\left(
\mathbf{X}^{\mathrm{T}}\mathbf{X}
+
\lambda I
\right)^{-1}
\mathbf{X}^{\mathrm{T}}\mathbf{y},
\end{align}
where \(\lambda\) denotes the regularization strength and \(I\) is the identity matrix.

In \ac{qrc}, we can alternatively make use of sampling noise as an effective regularizer by perturbing the feature matrix according to
\begin{align}
\mathbf{X}
\leftarrow
\mathbf{X}+\mu\mathcal{N},
\end{align}
which has been shown to be equivalent to Tikhonov regularization under appropriate conditions~\cite{BIS95}. Here, \(\mu=10^{-5}\) denotes the noise amplitude, while the entries of \(\mathcal{N}\) are sampled from a Gaussian distribution. The remaining \ac{qrc} simulation parameters are summarized in \cref{tab:qrc_parameters}.

\begin{table}[t]
\centering
\caption{Parameters used for the quantum reservoir computing simulations.}
\label{tab:qrc_parameters}
\begin{tabular}{ll}
\hline
Parameter & Setting \\
\hline
Training samples
    & $50\,000$ \\
Testing samples
    & $50\,000$ \\
\# of Hamiltonian
    & $10$ \\
Data re-uploading
    & $n_{\mathrm{enc}}\in\{1,\ldots,30\}$ \\
Temporal multiplexing
    & $N_V=30$ \\
\# of reservoir qubits
    & $L=4$ \\
Noise level
    & $\eta=10^{-5}$ \\
\hline
\end{tabular}
\end{table}

\section{Information Processing Capacity} \label{app:ipc}

\begin{definition}[Memory Capacity]
\label{def:memory_cap}
Given a uniformly distributed input sequence $\{u_n\} \in \mathcal{U}([-1,1])$, the system is trained to reconstruct previous inputs delayed by $d$ time steps, resulting in the target functions
$
y_n(d) = u_{n-d}, 
$
and predictions $\hat{y}_n(d)$. We denote the vectors corresponding to the full time series of length $N_T$ by $\mathbf{y}_d$ and $\hat{\mathbf{y}}_d$. 
The corresponding delay-dependent memory capacity is defined as
\begin{align}
    \mathrm{MC}(d)
    :=r^2(\mathbf{y}_d,\mathbf{y}_d)=
    \frac{\mathrm{cov}^2\!\left(\hat{\mathbf{y}}_d,\mathbf{y}_d\right)}
    {\sigma^2(\hat{\mathbf{y}}_d)\sigma^2(\mathbf{y}_d)},
    \label{eq:capacity}
\end{align}
where $\sigma^2(\cdot)$ denotes the variance and $r$ the Pearson correlation function \cite{JAE01}. 
The quantity $\mathrm{MC}(d)$ therefore measures how accurately the reservoir can reconstruct information from $d$ time steps in the past.
To characterize the total linear memory stored in the reservoir, we sum over all delays and obtain the total memory capacity, also referred to as the first-order information processing capacity,
$
    \mathrm{IPC}_1
    =
    \sum_d \mathrm{MC}(d).
$
\end{definition}

\begin{definition}[Fading Memory Property]  
\label{def:fading_mem}
A reservoir posses the fading memory property (\ac{fmp}), if 
\begin{align}
    \lim_{d\rightarrow \infty} 
    \mathrm{MC}(d)=0
\end{align}  
\end{definition}

While the memory capacity gives insight about memory of the system, it does not account for non-linear transformations of previous inputs, i.e. of the form $y_n(d) = u^2_{n-d}$ or $y_n(d_1,d_2)= u_{n-d_1}u^2_{n-d_2}$. To get a complete basis, it would be required that all target functions are othogonal under some chosen scalar product, allowing the individual capacities to be summed consistently according to \cref{eq:capacity}. \cite{DAM12} proposes exactly such a framework, where uniformly distributed inputs are considered and the notion of memory capacity is extended to nonlinear terms, referred to as the \textit{information processing capacity}.

\begin{definition}[Information processing capacity]
Let $(u_n)_{n\in\mathbb Z}$ be an input sequence with
$
u_n \sim \mathcal U([-1,1]),
$
and let $(l_k)_{k\ge0}$ denote the Legendre polynomials defined recursively with $l_0(x) = 1$ and $l_1(x) = x$ by
\begin{align}
    l_k(x)
    &=
    \frac{2k-1}{k}x\,l_{k-1}(x)
    -
    \frac{k-1}{k}l_{k-2}(x),
    \quad k\ge2.
\end{align}
The Legendre polynomials form an orthogonal basis on $[-1,1]$ with respect to the scalar product
$
    \langle f,g\rangle
    :=
    \int_{-1}^{1}
    f(x)g(x)\,\mathrm d x .
$
For delays $d_1,\ldots,d_n\in\mathbb N$ and polynomial orders
$k_1,\ldots,k_n\in\mathbb N_0$, define the target function
\begin{align}
    y_n^{(k_1,\ldots,k_m)}
    (d_1,\ldots,d_m)
    :=
    \prod_{i=1}^{m}
    l_{k_i}(u_{n-d_i}).
\end{align}

Let $\hat y_n^{(k_1,\ldots,k_m)}(d_1,\ldots,d_m)$ denote the corresponding prediction generated by the reservoir. The associated capacity is defined by
\[
    C\!\left(
        y,
        \hat y
    \right)
    :=
    \left(
        \frac{
            \operatorname{cov}(y,\hat y)
        }{
            \sigma(y)\sigma(\hat y)
        }
    \right)^2.
\]
The information processing capacity of order $D$ is obtained by summing the capacities over all target functions of total degree $D$,
\[
    \mathrm{IPC}_D
    :=
    \sum_{\sum_i k_i = D}
    C(y,\hat y).
\]
The total information processing capacity is then defined as
\begin{align}
    \mathrm{IPC}
    :=
    \sum_{D\ge1}
    \mathrm{IPC}_D .
\end{align}
\end{definition}

\section{Benchmark Tasks}\label{app:tasks}
Reservoir computing is typically applied to time-series prediction tasks. While the previous discussion provides a useful visualization of how the data re-uploading scheme influences the nonlinear response of the system, the spiral classification problem should be regarded primarily as a toy task. More realistic benchmark tasks in reservoir computing involve the prediction of chaotic dynamics. In this work, we therefore consider the Lorenz--63 system and the Mackey--Glass prediction task.

\textit{Mackey--Glass Task:} 
The first task we consider is the prediction of the Mackey--Glass system \cite{MAC77}, described by the equation
\begin{equation}
\frac{dx}{dt} 
=
\beta \frac{x(t - \tau_{MG})}{1 + x(t - \tau_{MG})^{10}}
-
\gamma x(t),
\end{equation}
with $\beta = 0.2$, $\gamma = 0.1$, and $\tau_{MG} = 17$. 
The system exhibits chaotic dynamics for $\tau_{MG} > 17$. 

The dynamics are discretized with time step $\Delta t = 1$, resulting in the discrete sequence
\[
x_n = x(n\Delta t),
\]
while the numerical integration is performed using $dt = 0.1$. 
The standard Mackey--Glass prediction task consists of predicting the future value $x_{n+10}$ from the current input $x_n$.

\textit{Lorenz-Task:}  
Another common benchmark task for nonlinear time-series prediction is the Lorenz--63 system~\cite{LOR63}. 
For the parameters $\sigma = 10$, $\rho = 28$, and $\beta = 8/3$, the time evolution is governed by
\begin{align}
    \dot{x} &= \sigma (y - x),\qquad \dot{y} &= x(\rho - z) - y, ,\qquad \dot{z} &= xy - \beta z,
    \label{eq:lorenz}
\end{align}
The system is discretized with time step $\Delta t = 0.1$ to obtain the discrete time series
\(
x_n = x(n\Delta t),~
y_n = y(n\Delta t),~
z_n = z(n\Delta t),
\)
where the numerical integration is performed using $dt = 0.001$.  We consider a setting in which the reservoir only receives access to the signal $x_n$. The first task is to predict the future value $x_{n+1}$, which we denote as the Lorenz \(x \rightarrow x\) task. The second task is to predict the corresponding value $z_n$ from the input $x_n$, which we denote as the Lorenz \(x \rightarrow z\) task.

\begin{figure}[H]
    \centering
    \includegraphics[scale=1]{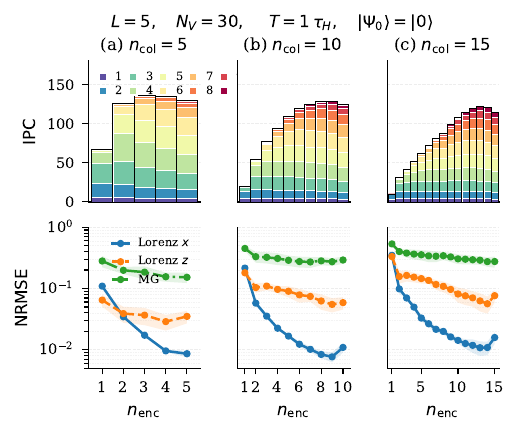}
    \caption{
    IPC (first row) and NRMSE (second row) for the hybrid protocol with a fixed number of collisions $n_{\mathrm{col}}$. The left limit corresponds to the dissipative protocol, while the right limit corresponds to the data re-uploading protocol.
    }
    \label{fig:app_hybrid}
\end{figure}
\section{Hybrid Algorithm Results}\label{sec:hybrid}

\Cref{fig:app_hybrid} shows the task performance for different fixed numbers of collisions $n_{\mathrm{col}}$. For few collisions, the data re-uploading protocol performs best across all tasks. As the number of collisions increases, performance improves near the data re-uploading limit, with optimal results at $(n_{\mathrm{enc}}, n_{\mathrm{col}}) = (10, 11)$ and $(14, 15)$. Furthermore, the hybrid protocol exhibits local maxima in IPC across all three realizations. Further results are given in Supplemental Material \cite{Supp}, Sec. III.

%%%%%%%%%%%%%%%%%%%%%%%%%%%%%%%%%%%%%%%
%%%%%%%%%%%%%% Hybrid Optimal Configurations %%%%%%%%%%%
%%%%%%%%%%%%%%%%%%%%%%%%%%%%%%%%%%%%%%%%%%%%%%%%%%%%%%%%

\begin{table}[H]
    \centering
    \caption{
    Optimal hybrid configurations for $L=5$, $N_v=30$,
    and $T=T_H$. The IPC values are maximized, while the
    prediction errors (NRMSE) are minimized.
    Collision and encoding counts include the initial encoding.
    }
    \label{tab:hybrid_optima}
    \begin{tabular}{lccc}
        \hline\hline
        Task
        & Optimal value
        & $n_{\mathrm{col}}$
        & $n_{\mathrm{enc}}$ \\
        \hline
        Linear IPC
        & $9.486$
        & $1$ & $1$ \\

        Nonlinear IPC
        & $135.181$
        & $2$ & $1$ \\

        Total IPC
        & $143.368$
        & $2$ & $1$ \\

        Mackey--Glass NRMSE
        & $0.01293$
        & $1$ & $1$ \\

        Lorenz $x$ NRMSE
        & $0.00664$
        & $7$ & $6$ \\

        Lorenz $z$ NRMSE
        & $0.01377$
        & $2$ & $2$ \\
        \hline\hline
    \end{tabular}
\end{table}

\bibliography{lit}

\end{document}

\title{Supplemental Material for\\
``Data Reuploading in Quantum Reservoir Computing''}

\author{Saud \v{C}indrak}
\ILMENAU

\author{Kaito Kobayashi}
\UTOKYO

\author{Lara Giebeler}
\ITPTUB

\author{Kathy L\"{u}dge}
\email{kathy.luedge@tu-ilmenau.de}
\ILMENAU

\date{\today}

\maketitle

%%%%%%%%%%%%%%%%%%%%%%%%%%%%%%%%%%%%%%%%%%%%%%%%%%%%%%%%%%%%
% Supplemental numbering
%%%%%%%%%%%%%%%%%%%%%%%%%%%%%%%%%%%%%%%%%%%%%%%%%%%%%%%%%%%%

\renewcommand{\theequation}{S\arabic{equation}}
\renewcommand{\thefigure}{S\arabic{figure}}
\renewcommand{\thetable}{S\arabic{table}}

%%%%%%%%%%%%%%%%%%%%%%%%%%%%%%%%%%%%%%%%%%%%%%%%%%%%%%%%%%%%
% Supplement
%%%%%%%%%%%%%%%%%%%%%%%%%%%%%%%%%%%%%%%%%%%%%%%%%%%%%%%%%%%%

\section{Memory--Nonlinearity Tradeoff via Data Re-uploading}
\label{sup:mc_increase}

\begin{figure}[htbp]
  \centering
  \includegraphics[scale = 1]
    {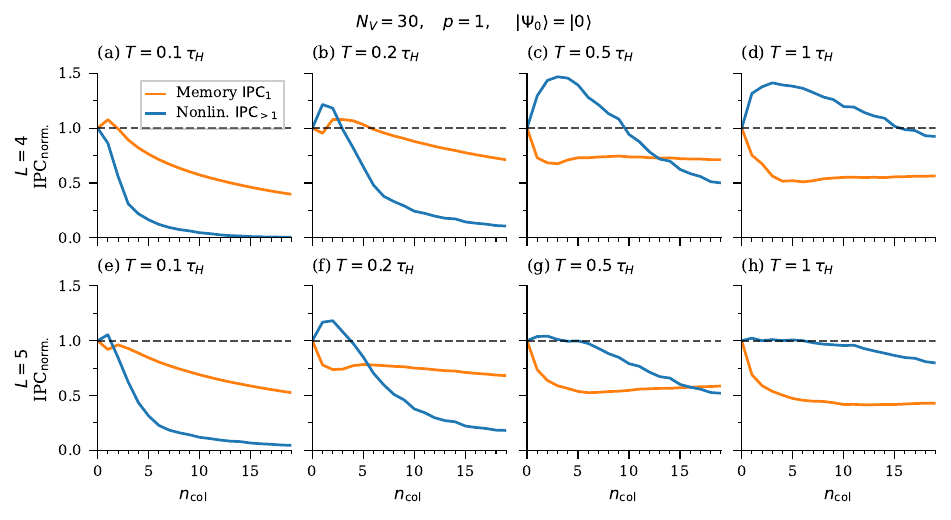}
  \caption{
    Linear memory contribution \(\mathrm{IPC}_1\) and
    nonlinear contribution
    \(\mathrm{IPC}_{>1}=\sum_{n>1}\mathrm{IPC}_n\)
    under data re-uploading.
    Panels (a--d) show the results for the \(L=4\)
    reservoir at clock cycles \(T/\tau_H=0.1\),
    \(0.2\), \(0.5\), and \(1.0\), respectively,
    while panels (e--h) show the corresponding
    results for \(L=5\).
  }
  \label{fig:qrc_mem_nonlin}
\end{figure}

% Add further supplemental content here.

Here, we investigate how data re-uploading affects the balance between linear memory and nonlinear information processing for reservoirs of \(L=4\) and \(L=5\) qubits. We quantify the linear contribution by the first-order information processing capacity, \(\mathrm{IPC}_1\), while the nonlinear contribution is defined as the sum of all higher-order capacities,

$$
    \mathrm{IPC}_{>1} = \sum_{n>1} \mathrm{IPC}_n .
$$

The corresponding results are shown in \cref{fig:qrc_mem_nonlin}, with the \(L=4\) results in the first row and the \(L=5\) results in the second row.
For the shortest clock cycle, \(T=0.1\tau_H\), both the linear and nonlinear contributions decrease as the number of data re-uploading steps is increased. At \(T=0.2\tau_H\), a different behavior begins to emerge: the nonlinear contribution increases with additional re-uploading, while the linear contribution changes only weakly for \(L=4\) and decreases for \(L=5\).
This tradeoff becomes considerably more pronounced for the longer clock cycles, \(T=0.5\tau_H\) and \(T=1.0\tau_H\). In these regimes, data re-uploading substantially enhances the higher-order information processing capacity compared with the vanilla QRC protocol, while the first-order contribution generally decreases. Thus, the additional encoding operations increasingly redistribute the information processing capacity from linear memory toward nonlinear processing as the clock cycle becomes longer.

For \(L=5\), the dependence at small numbers of re-uploading steps is somewhat weaker, resulting in an approximately flat initial behavior for some clock cycles. Nevertheless, the same overall trend is observed, as for \(T=0.5\tau_H\) we additionally find a small initial enhancement of the nonlinearity by a following decrease. This lack of enhancement of overall nonlinearity might be due to only showing results only up to the tenth order.  

Consequently, the reported value of the nonlinearity\(IPC_{>1}\) may underestimate the total nonlinear contribution. %The observed redistribution toward higher-order IPCs therefore provides further evidence that data re-uploading enhances the nonlinear processing capabilities of the reservoir at the expense of linear memory.

Taken together, these results with the results discussed in the main text show that an inherent memory-nonlinearity tradeoff due to data reuploading.

 \section{Dependence on the initial quantum state}
\label{sec:initial_state}

%Here, we investigate whether the performance of the dissipative protocol depends on the choice of the initial state of the quantum reservoir. 

In the dissipative protocol, instead of re-encoding the input \(u_k\) at the corresponding encoding intervals, the designated qubit is repeatedly reset to a fixed state \(\ket{\Psi_0}\). To verify that the observed behavior is not a consequence of specifically resetting the qubit to \(\ket{\Psi_0}=\ket{0}\), as considered in the main text, we repeat the simulations for three different reset states:
\(\ket{\Psi_0}=\ket{0}\) (first row),
\(\ket{\Psi_0}=\ket{+}\) (second row), and
\(\ket{\Psi_0}=\ket{\Psi_{\mathrm{rand}}}\) (third row).
Here, \(\ket{\Psi_{\mathrm{rand}}}\) denotes a randomly generated single-qubit state obtained.

\begin{figure}[t]
\centering
\includegraphics[scale=0.85]{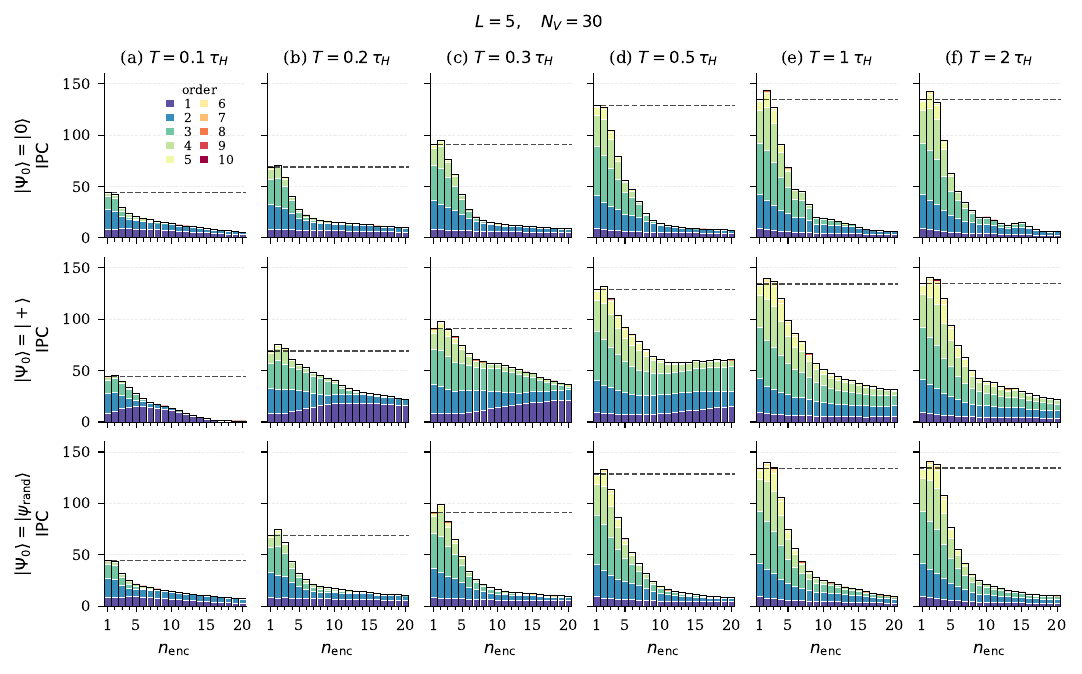}
\caption{
Information processing capacity for the dissipative protocol with different reset states of the quantum reservoir for \(L=5\) qubits.
}
\label{fig:suppl_state_L5}
\end{figure}

The corresponding results for reservoirs of \(L=4\) and \(L=5\) qubits are shown in \cref{fig:suppl_state_L4,fig:suppl_state_L5}, respectively. Overall, the qualitative behavior is largely insensitive to the chosen reset state. For a small number of encoding operations, the information processing capacity exhibits a modest enhancement relative to the vanilla QRC protocol with \(n_{\mathrm{enc}}=1\), followed by a gradual decrease as the number of encodings is increased. Although the magnitude of the initial enhancement varies only slightly between the different reset states, some quantitative differences are visible. In particular, for the \(\ket{+}\) reset state, the decay appears to occur more slowly, resulting in a more extended tail that approaches a plateau at larger numbers of encodings. For larger clock cycles $T$, the state has more time to spread throughout the Hilbert space and thus become more independent of the encoded state. Additionally, the falloff is due to the repeated encoding of the first qubit, which restricts the time evolution and can result in partial freezing, thereby preventing information from the encoding qubit from propagating to the rest of the system.

These results show that the dissipative protocol is not tied to a particular choice of reset state. For both \(L=4\) and \(L=5\), we consistently observe a small enhancement over the vanilla QRC performance at low \(n_{\mathrm{enc}}\), followed by a reduction in capacity as the number of reset operations increases. The dependence on the specific initial state therefore appears to be predominantly quantitative.

%%%%%%%%%%%%%%%%%%%%%%%%%%%%%%%%%%%%%%%%%%%%%%%%%%%%%%%%%%
%%%%%%%%%%%%%%%%%%%%%%%%%%%%%%%%%%%%%%%%%%%%%%%%%%%%%%%%%%
%%%%%%%%%%%%%%%%%%%%%%%%%%%%%%%%%%%%%%%%%%%%%%%%%%%%%%%%%%
%%%%%%%%%%%%%%%%%%%%%%%%%%%%%%%%%%%%%%%%%%%%%%%%%%%%%%%%%%
%%%%%%%%%%%%%%%%%%%%%%%%%%%%%%%%%%%%%%%%%%%%%%%%%%%%%%%%%%

\section{Interleaved Hybrid Protocol}
\label{sec:hybrid_fixed}

\begin{figure}[H]
    \centering

    \begin{quantikz}[
        row sep={0.5cm,between origins},
        column sep=0.35cm
    ]
        \lstick{$q_1$}
        & \gate{\ket{\psi_{\mathrm{enc}}(u_k)}}
        & \gate[5]{U(\Delta t)}
        & \gate{\ket{\Psi_0}} \wireoverride{n}
        & \gate[5]{U(\Delta t)}
        & \gate{\ket{\psi_{\mathrm{enc}}(u_k)}} \wireoverride{n}
        & \gate[5]{U(\Delta t)}
        & \gate{\ket{\Psi_0}} \wireoverride{n}
        & \gate[5]{U(\Delta t)}
        & \qw
        \\
        \lstick{$q_2$}
        & \qw & & \qw & & \qw & & \qw & & \qw
        \\
        \lstick{$q_3$}
        & \qw & & \qw & & \qw & & \qw & & \qw
        \\
        \lstick{$q_4$}
        & \qw & & \qw & & \qw & & \qw & & \qw
        \\
        \lstick{$q_5$}
        & \qw & & \qw & & \qw & & \qw & & \qw
    \end{quantikz}

    \caption{
        Quantum circuit illustrating the deterministic hybrid
        protocol for a five-qubit reservoir with
        $n_{\mathrm{col}}=4$ collisions.
        At each collision, the first reservoir qubit is
        overwritten with either the input-dependent state
        $\ket{\psi_{\mathrm{enc}}(u_k)}$
        or the fixed reset state $\ket{\Psi_0}$.
        The disconnected wire segments indicate that the
        previous state of the first qubit is discarded.
        Between collisions, the reservoir evolves under
        $U(\Delta t)=e^{-iH\Delta t}$,
        with $\Delta t=T/n_{\mathrm{col}}$.
        The illustrated sequence corresponds to
        $n_{\mathrm{enc}}=2$ and $n_{\mathrm{reset}}=2$.
    }
    \label{fig:hybrid_deterministic_circuit}
\end{figure}
 
We introduce a deterministic hybrid protocol that combines data re-uploading and dissipative state resets within a
single reservoir evolution cycle. For this purpose, we consider
\begin{align}
    n_{\mathrm{col}}
    =
    n_{\mathrm{enc}} + n_{\mathrm{reset}}
\end{align}
state overwrites, where $n_{\mathrm{enc}}$ denotes the number of input encodings and $n_{\mathrm{reset}}$ the number of dissipative state resets.

The collisions occur at equally spaced times $t_j=jT/n_{\mathrm{col}}$, with $j=0,\ldots,n_{\mathrm{col}}-1$ and $\Delta t=T/n_{\mathrm{col}}$. The initial collision at $t_0=0$ always encodes the input $u_k$, while subsequent collisions either re-encode the same input or reset the first reservoir qubit to a fixed state.
The input-dependent encoded state is
\begin{align}
    \ket{\psi_{\mathrm{enc}}(u_k)}
    =
    \sqrt{\frac{1-u_k}{2}}\,\ket{0}
    +
    \sqrt{\frac{1+u_k}{2}}\,\ket{1},
    \label{eq:hybrid_fixed_encoding}
\end{align}
while the fixed reset state is chosen as $\ket{\Psi_0}=\ket{00000}$.
The state prepared at the $j$-th collision of the $k$-th input is
\begin{align}
    \ket{\psi_{k,j}^{(1)}}
    =
    \begin{cases}
        \ket{\psi_{\mathrm{enc}}(u_k)},
        & j\in\mathcal{I}_{\mathrm{enc}},\\[2mm]
        \ket{\Psi_0},
        & j\in\mathcal{I}_{\mathrm{reset}},
    \end{cases}
    \label{eq:hybrid_fixed_state}
\end{align}
where $\mathcal{I}_{\mathrm{enc}}$ and $\mathcal{I}_{\mathrm{reset}}$ denote the sets of encoding and reset collision indices, respectively.

The reservoir update corresponding to the
$j$-th collision is then given by
\begin{align}
    \mathcal{G}_{\Delta t}^{(k,j)}(\rho)
    &=
    e^{-iH\Delta t}
    \left(
    \ket{\psi_{k,j}^{(1)}}
    \bra{\psi_{k,j}^{(1)}}
    \otimes
    \operatorname{Tr}_1[\rho]
    \right)
    e^{iH\Delta t}.
    \label{eq:hybrid_fixed_map}
\end{align}
Here, the state of the first reservoir qubit is overwritten at each collision, followed by coherent evolution under the reservoir Hamiltonian $H$. 

The complete evolution over one input cycle is obtained by composing the individual
collision maps,
\begin{align}
    \rho_k
    &=
    \mathcal{G}_{\Delta t}^{(k,n_{\mathrm{col}}-1)}
    \circ \cdots \circ
    \mathcal{G}_{\Delta t}^{(k,0)}
    (\rho_{k-1}).
    \label{eq:hybrid_fixed_full_map}
\end{align}

Figure~\ref{fig:hybrid_deterministic_circuit} illustrates the protocol for a five-qubit reservoir with $n_{\mathrm{col}}=4$ collisions with $n_{\mathrm{enc}}=n_{\mathrm{col}}=2$. 
For a fixed number of collisions, $n_{\mathrm{col}}$, varying the number of encodings $n_{\mathrm{enc}}$ allows us to investigate the interplay between data re-uploading and dissipation. The additional encodings are distributed approximately uniformly over the available collision times, while all remaining collisions perform qubit resets.

\newpage
\subsection{Results for the Interleaved Hybrid Protocol}
\begin{figure}[H]
    \centering
    \begin{subfigure}[t]{\textwidth}
        \centering
        \includegraphics[scale = 1]{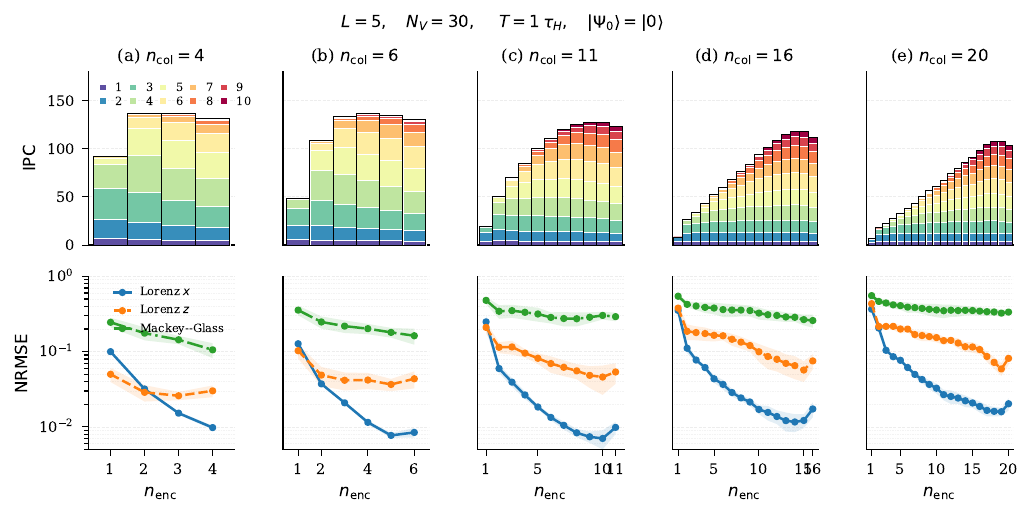}
        %\caption{$T=\tau_H$.}
        \label{fig:hybrid_fixed_L5_T1}
    \end{subfigure}

    \vspace{0.4cm}

    \begin{subfigure}[t]{\textwidth}
        \centering
        \includegraphics[scale = 1]{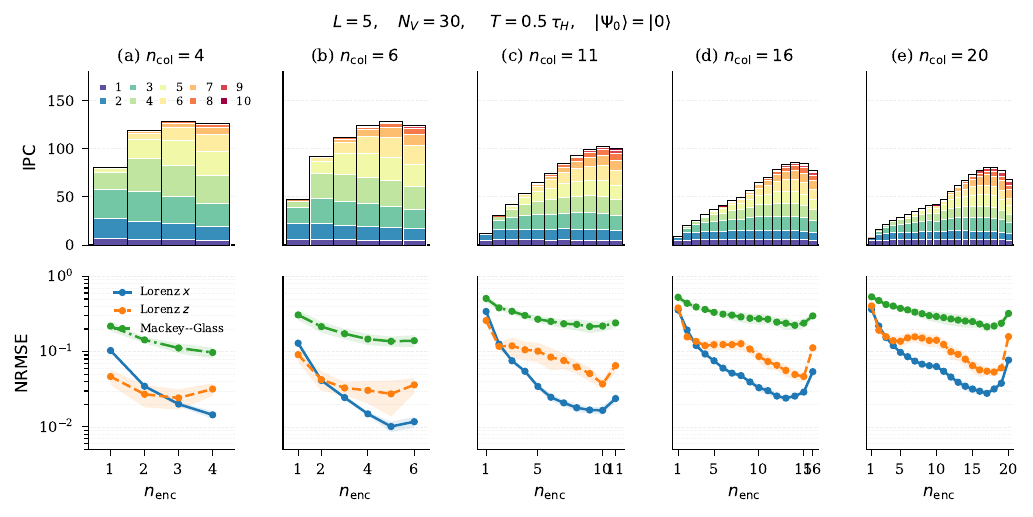}
        %\caption{$T=0.5\tau_H$.}
        \label{fig:hybrid_fixed_L5_T05}
    \end{subfigure}

    \caption{
        Information processing capacity (IPC) and normalized root mean squared error (NRMSE) of the deterministic hybrid protocol for a five-qubit reservoir ($L=5$) with $N_V=30$. The columns correspond to different numbers of collisions $n_{\mathrm{col}}\in\{4,6,11,16,20\}$, while the number of input encodings $n_{\mathrm{enc}}$ is varied. The upper panels show the IPC contributions up to tenth order, and the lower panels show the prediction errors for Lorenz $x$, Lorenz $z$, and Mackey--Glass. Results are shown for  $T=\tau_H$ and $T=0.5\tau_H$. Intermediate configurations combine data re-uploading and dissipative resets.
    }
    \label{fig:hybrid_fixed_L5}
\end{figure}

We investigate the information processing capacity (IPC) and task performance of the deterministic hybrid protocol for different numbers of collisions $n_{\mathrm{col}}$ and input encodings $n_{\mathrm{enc}}$. Figure~\ref{fig:hybrid_fixed_L5} shows the results for a five-qubit reservoir with $n_{\mathrm{col}}\in{4,6,11,16,20}$ and evolution times $T=\tau_H$ and $T=0.5\tau_H$.

For $T=\tau_H$, the total IPC generally increases with the number of encodings, reaching its maximum for hybrid configurations close to the pure data re-uploading limit. The prediction tasks, however, exhibit a more differentiated behavior. For small numbers of collisions, only the Lorenz $z$ task benefits from combining encoding and resetting. Already at $n_{\mathrm{col}}=6$, a local minimum in the prediction error is observed at $n_{\mathrm{enc}}=5$, indicating that the hybrid protocol slightly outperforms pure data re-uploading. As the number of collisions increases [Panel (c-e)], this advantage becomes more pronounced, with both Lorenz prediction tasks exhibiting lower prediction errors for hybrid configurations than for pure data re-uploading.

For $T=0.5\tau_H$, similar trends are observed. Increasing the number of encodings generally enhances nonlinear information processing, while the optimal prediction performance depends on the task and collision configuration. The highest total IPC is again achieved for hybrid configurations close to the data re-uploading limit. For larger numbers of collisions, particularly $n_{\mathrm{col}}\in{11,16,20}$, pronounced minima in the prediction errors emerge at intermediate values of $n_{\mathrm{enc}}$. In this regime, both Lorenz prediction tasks and the Mackey--Glass task can benefit from hybrid configurations containing only a small number of dissipative resets.

The corresponding results for a four-qubit reservoir are shown in Fig.~\ref{fig:hybrid_fixed_L4}. The qualitative behavior is similar, although the overall IPC is smaller. In particular, the task-dependent minima in the prediction errors persist, demonstrating that the beneficial interplay between data re-uploading and dissipation is not restricted to the five-qubit reservoir.

These results highlight the potential of the deterministic hybrid protocol to enhance computational performance beyond that achievable through either mechanism alone. By introducing a small number of dissipative resets into an otherwise data-re-uploading-dominated evolution, the reservoir can achieve improved prediction accuracy while retaining strong nonlinear information-processing capabilities.

%%%%%%%%%%%%%%%%%%%%%%%%%%%%%%%%%%%%%%%%%%%%%%%%%%%%%%%%%%
%%%%%%%%%%%%%%%%%%%%%%%%%%%%%%%%%%%%%%%%%%%%%%%%%%%%%%%%%%
%%%%%%%%%%%%%%%%%%%%%%%%%%%%%%%%%%%%%%%%%%%%%%%%%%%%%%%%%%
%%%%%%%%%%%%%%%%%%%%%%%%%%%%%%%%%%%%%%%%%%%%%%%%%%%%%%%%%%
%%%%%%%%%%%%%%%%%%%%%%%%%%%%%%%%%%%%%%%%%%%%%%%%%%%%%%%%%%

\section{Probabilistic Hybrid Protocol}
\label{sec:hybrid}

Another way to investigate the transition between the dissipative protocol and the data re-uploading protocol is to introduce a hybrid approach that continuously interpolates between the two. Recall that the reservoir update map is given by
\begin{align}
\rho_k
&=
\mathcal{G}_{T}(\rho_{k-1})
 =
e^{-iHT}
\left(
\ket{\psi_k^{(1)}}\bra{\psi_k^{(1)}}
\otimes
\operatorname{Tr}_1[\rho_{k-1}]
\right)
e^{iHT},
\label{eq:map_G}
\end{align}
where the state of the first qubit is replaced before each reservoir evolution step.

For the dissipative protocol, the first qubit is reset to a fixed state,
\begin{align}
\ket{\psi_k^{(1)}} = \ket{\Psi_0}.
\end{align}
In contrast, for the data re-uploading protocol, the first qubit is prepared
in the input-dependent state
\begin{align}
\ket{\psi_k^{(1)}}
&=
\ket{\psi_{\mathrm{enc}}(u_k)}=
\sqrt{\frac{1-u_k}{2}}\,\ket{0}
+
\sqrt{\frac{1+u_k}{2}}\,\ket{1},
\label{eq:encoding_state}
\end{align}
where \(u_k\) denotes the data point encoded at the \(k\)-th time step.

A continuous interpolation between the two protocols can be constructed by
introducing the mixed state
\begin{align}
\rho_{k,p}^{(\mathrm{hybrid})}
&=
p\,
\ket{\psi_{\mathrm{enc}}(u_k)}
\bra{\psi_{\mathrm{enc}}(u_k)}
+
(1-p)\,
\ket{\Psi_0}\bra{\Psi_0},
\qquad
p \in [0,1].
\label{eq:hybrid_state}
\end{align}
The parameter \(p\) controls the relative contribution of data re-uploading and dissipation. For \(p=0\), the hybrid protocol reduces to the purely dissipative protocol, whereas for \(p=1\), it reproduces the data re-uploading protocol.

The corresponding reservoir update map is therefore
\begin{align}
\rho_k
&=
\mathcal{G}_{T}^{(p)}(\rho_{k-1})
 =
e^{-iHT}
\left(
\rho_{k,p}^{(\mathrm{hybrid})}
\otimes
\operatorname{Tr}_1[\rho_{k-1}]
\right)
e^{iHT}.
\label{eq:map_G_hybrid}
\end{align}

This hybrid construction allows us to continuously tune between input-independent resetting and input-dependent data re-uploading, and thus to investigate how the reservoir dynamics and information-processing properties change as the influence of the encoded input is increased.

\subsection{Ancilla-assisted realization}

The mixed state in Eq.~\eqref{eq:hybrid_state} can be generated by introducing
an ancillary qubit and a corresponding qunatum circuit is shown in \cref{fig:hybrid_ancilla}. We denote the ancilla by \(q_1\) and the reservoir input
qubit by \(q_2\), and initialize the two-qubit system in
\begin{align}
\ket{q_1 q_2}_0 = \ket{00}.
\end{align}

The ancilla is first rotated according to
\begin{align}
R_y(\theta_p)\ket{0}
=
\sqrt{p}\,\ket{0}
+
\sqrt{1-p}\,\ket{1},
\end{align}
with
$
\theta_p
=
2\arccos\left(\sqrt{p}\right). 
$
After applying the rotation to the first qubit, the state becomes
\begin{align}
\ket{q_1 q_2}_1
&=
\left(
R_y(\theta_p)\otimes I
\right)
\ket{00}
=
\sqrt{p}\,\ket{00}
+
\sqrt{1-p}\,\ket{10}.
\end{align}

Next, an \(X\) gate is applied to the ancilla,
\begin{align}
\ket{q_1 q_2}_2
&=
\left(
X\otimes I
\right)
\ket{q_1 q_2}_1
 =
\sqrt{p}\,\ket{10}
+
\sqrt{1-p}\,\ket{00}.
\end{align}
This allows the encoded input state to be prepared on the second qubit using a conventional controlled operation that is activated when the ancilla is in the state \(\ket{1}\). We define \(U_{\mathrm{enc}}(u_k)\) such that
$
U_{\mathrm{enc}}(u_k)\ket{0}
=
\ket{\psi_{\mathrm{enc}}(u_k)}.
$
Applying the controlled operation \(C(U_{\mathrm{enc}}(u_k))\) therefore gives
\begin{align}
\ket{q_1 q_2}_3
&=
C(U_{\mathrm{enc}}(u_k))
\ket{q_1 q_2}_2
 =
\sqrt{p}\,
\ket{1}
\ket{\psi_{\mathrm{enc}}(u_k)}
+
\sqrt{1-p}\,
\ket{00}.
\end{align}

The ancilla is then flipped back,
\begin{align}
\ket{q_1 q_2}_4
&=
\left(
X\otimes I
\right)
\ket{q_1 q_2}_3
 =
\sqrt{p}\,
\ket{0}
\ket{\psi_{\mathrm{enc}}(u_k)}
+
\sqrt{1-p}\,
\ket{10}.
\end{align}

Finally, we apply a second controlled unitary \(C(U_{\Psi_0})\), where
$
U_{\Psi_0}\ket{0}
=
\ket{\Psi_0}.
$
This operation acts on the second qubit only when the ancilla is in
\(\ket{1}\), yielding
\begin{align}
\ket{q_1 q_2}_5
&=
C(U_{\Psi_0})
\ket{q_1 q_2}_4
 =
\sqrt{p}\,
\ket{0}
\ket{\psi_{\mathrm{enc}}(u_k)}
+
\sqrt{1-p}\,
\ket{1}
\ket{\Psi_0}.
\label{eq:hybrid_purification}
\end{align}

Equivalently, the complete circuit can be written as
\begin{align}
\ket{q_1 q_2}_5
&=
C(U_{\Psi_0})
\left(
X\otimes I
\right)
C(U_{\mathrm{enc}}(u_k))
\left(
X\otimes I
\right)
\left(
R_y(\theta_p)\otimes I
\right)
\ket{00}.
\label{eq:hybrid_circuit}
\end{align}

The state in Eq.~\eqref{eq:hybrid_purification} is a purification of the
hybrid mixed state. 
Using
\begin{align}
\operatorname{Tr}_1
\left[
\ket{i}\bra{j}\otimes A
\right]
=
\bra{j}\ket{i}\,A,
\end{align}
the off-diagonal terms in the ancilla basis vanish due to \(\bra{0}\ket{1}=\bra{1}\ket{0}=0\), while the diagonal terms remain.

The corresponding density operator is
\begin{align}
\ket{q_1 q_2}_5\bra{q_1 q_2}_5
=&
p\,
\ket{0}\bra{0}
\otimes
\ket{\psi_{\mathrm{enc}}(u_k)}
\bra{\psi_{\mathrm{enc}}(u_k)}
+
\sqrt{p(1-p)}\,
\ket{0}\bra{1}
\otimes
\ket{\psi_{\mathrm{enc}}(u_k)}
\bra{\Psi_0}\nonumber \\
&+
\sqrt{p(1-p)}\,
\ket{1}\bra{0}
\otimes
\ket{\Psi_0}
\bra{\psi_{\mathrm{enc}}(u_k)}
+
(1-p)\,
\ket{1}\bra{1}
\otimes
\ket{\Psi_0}\bra{\Psi_0}.
\end{align}

Tracing out the ancillary qubit \(q_1\) removes the off-diagonal terms in the ancilla basis and gives
\begin{align}
\rho_{k,p}^{(\mathrm{hybrid})}
&=
\operatorname{Tr}_{1}
\left[
\ket{q_1 q_2}_5\bra{q_1 q_2}_5
\right]
\nonumber\\
&=
p\,
\ket{\psi_{\mathrm{enc}}(u_k)}
\bra{\psi_{\mathrm{enc}}(u_k)}
+
(1-p)\,
\ket{\Psi_0}\bra{\Psi_0}.
\label{eq:hybrid_state_from_ancilla}
\end{align}

Thus, the desired hybrid state can be generated coherently by coupling the reservoir input qubit to an ancilla and subsequently tracing out the ancilla. The parameter \(p\) is controlled entirely by the rotation angle \(\theta_p\), providing a continuous interpolation between the dissipative and data re-uploading limits.

\begin{figure}[t]
    \centering

     \begin{quantikz}[
    row sep=0.6cm,
    column sep=0.5cm, 
]
    \lstick{$\ket{0}_{q_1}$}
    & \slice{$\ket{q_1q_2}_0$}
    & \gate{R_y(\theta_p)}
    & \slice{$\ket{q_1q_2}_1$}
    & \gate{X}
    & \slice{$\ket{q_1q_2}_2$}
    & \ctrl{1}
    & \slice{$\ket{q_1q_2}_3$}
    & \gate{X}
    & \slice{$\ket{q_1q_2}_4$}
    & \ctrl{1}
    & \slice{$\ket{q_1q_2}_5$}
    & \qw
    \\
    \lstick{$\ket{0}_{q_2}$}
    & \qw
    & \qw
    & \qw
    & \qw
    & \qw
    & \gate{U_{\mathrm{enc}}(u_k)}
    & \qw
    & \qw
    & \qw
    & \gate{U_{\Psi_0}}
    & \qw
    & \qw
\end{quantikz}
    \caption{
        Ancilla-assisted realization of the hybrid input state. The ancilla \(q_1\) is prepared in a \(p\)-dependent superposition using \(R_y(\theta_p)\). Controlled operations subsequently prepare the reservoir input qubit \(q_2\) either in the encoded state \(\ket{\psi_{\mathrm{enc}}(u_k)}\) or in the fixed reset state \(\ket{\Psi_0}\). Tracing out the ancilla yields the mixed hybrid state \(\rho_{k,p}^{(\mathrm{hybrid})}\).
    }
    \label{fig:hybrid_ancilla}
\end{figure}

\subsection{Results for the Probabilistic Hybrid Algorithm}

\begin{figure}[t]
    \centering

    \begin{subfigure}[t]{0.48\textwidth}
        \centering
        \includegraphics[width=\linewidth]{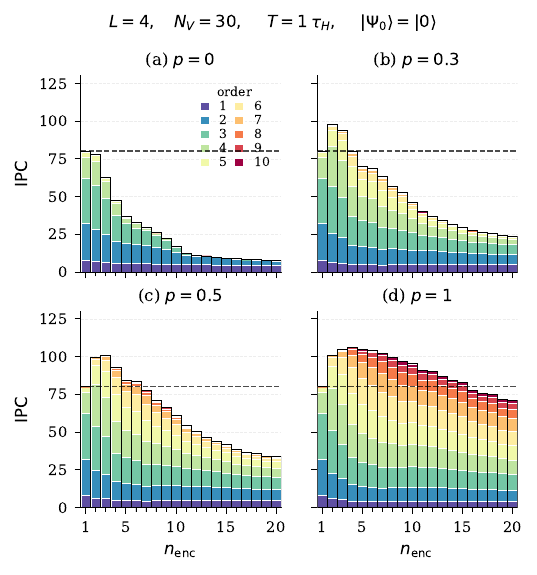}
        \caption{Hybrid protocol for a reservoir of \(L=4\) qubits.}
        \label{fig:fig3_L4}
    \end{subfigure}
    \hfill
    \begin{subfigure}[t]{0.48\textwidth}
        \centering
        \includegraphics[width=\linewidth]{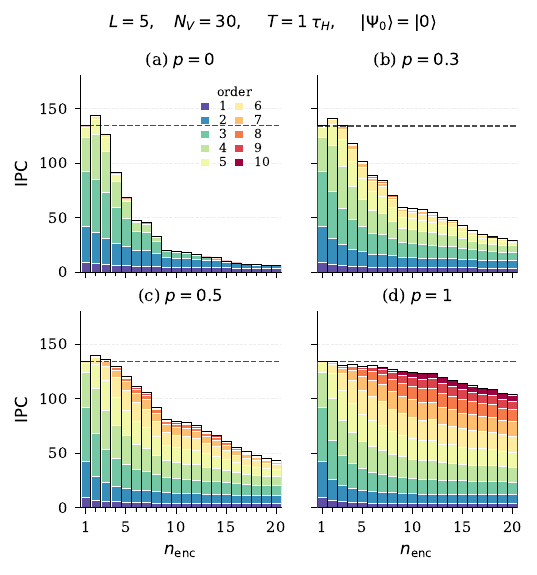}
        \caption{Hybrid protocol for a reservoir of \(L=5\) qubits.}
        \label{fig:fig3_L5}
    \end{subfigure}

    \caption{
        Information processing capacity up to tenth order as a function of the
        number of encodings \(n_{\mathrm{enc}}\) for the hybrid protocol.
        The parameter \(p\) interpolates between the dissipative and data
        re-uploading protocols: \(p=0\) corresponds to the dissipative protocol,
        \(p=0.3\) and \(p=0.5\) to intermediate hybrid protocols, and \(p=1\)
        to the data re-uploading protocol.
    }
    \label{fig:fig3_hybrid}
\end{figure}

We now investigate the expressivity of the hybrid protocol for reservoirs of \(L=4\) and \(L=5\) qubits. The corresponding results are shown in \cref{fig:fig3_L4,fig:fig3_L5}. The parameter \(p\) continuously interpolates between the dissipative and data re-uploading limits: \(p=0\) corresponds to the purely dissipative protocol, \(p=1\) to the full data re-uploading protocol, while \(p=0.3\) and \(p=0.5\) represent intermediate realizations of the hybrid scheme.

For both system sizes, the dissipative protocol (\(p=0\)) exhibits the weakest higher-order information processing capacities. As \(p\) is increased, the contributions from higher-order IPCs become progressively more pronounced. The intermediate values \(p=0.3\) and \(p=0.5\) already show a clear enhancement of nonlinear information processing relative to the purely dissative case, while the strongest nonlinear contributions are observed in the data re-uploading limit \(p=1\).

The same qualitative trend is present for both \(L=4\) and \(L=5\), indicating that the enhancement is not specific to a particular reservoir size. These results further support the interpretation that repeated data re-uploading acts as an important resource for increasing the nonlinear processing capabilities of the quantum reservoir.

\begin{figure}[t]
    \centering
    \includegraphics[scale=1]{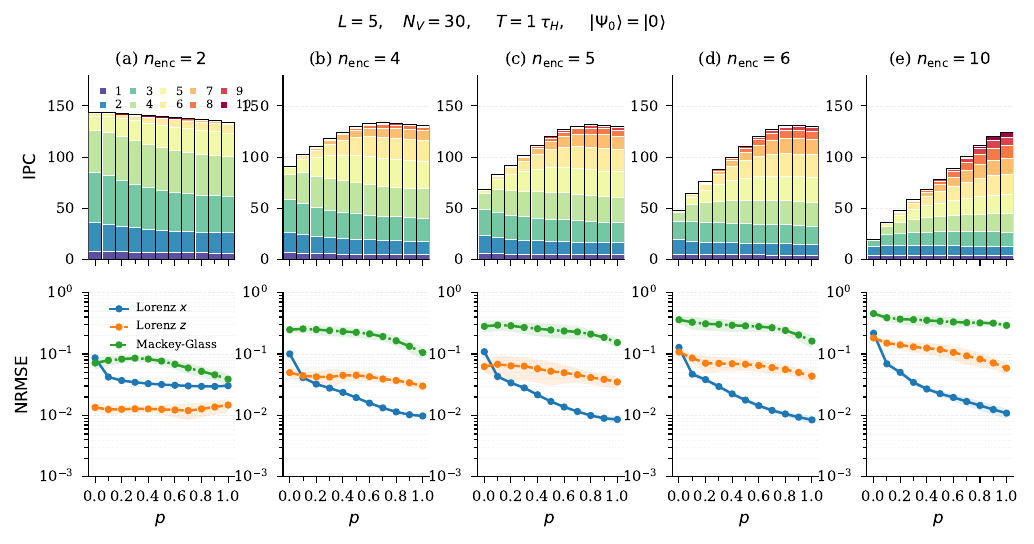}
    \caption{
        Dependence of the total information processing capacity and task
        performance on the hybrid interpolation parameter \(p\) for a reservoir
        of \(L=5\) qubits. Results are shown for
        \(n_{\mathrm{enc}} \in \{2,4,5,6,10\}\).
        The limits \(p=0\) and \(p=1\) correspond to the dissipative and
        data re-uploading protocols, respectively, while intermediate values
        describe the hybrid protocol.
    }
    \label{fig:fig5_L5}
\end{figure}

\begin{figure}[H]
\centering
\includegraphics[scale=0.75]{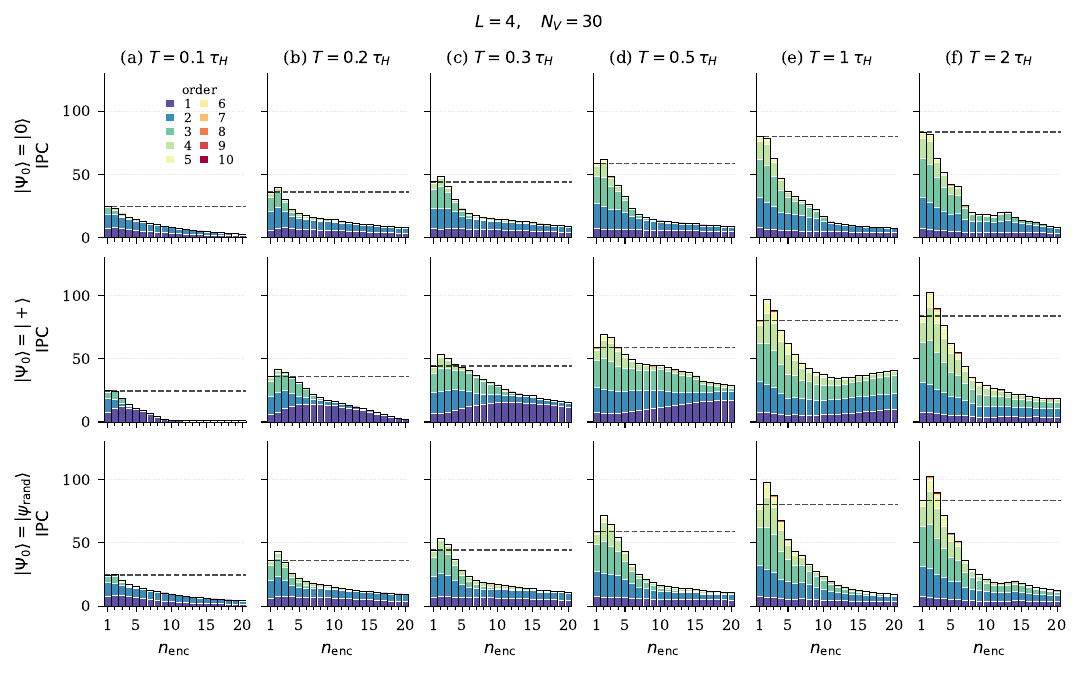}
\caption{
Information processing capacity for the dissipative protocol with different reset states of the quantum reservoir for \(L=4\) qubits.
}
\label{fig:suppl_state_L4}
\end{figure}

\begin{figure}[H]
    \centering
    \includegraphics[scale=1]{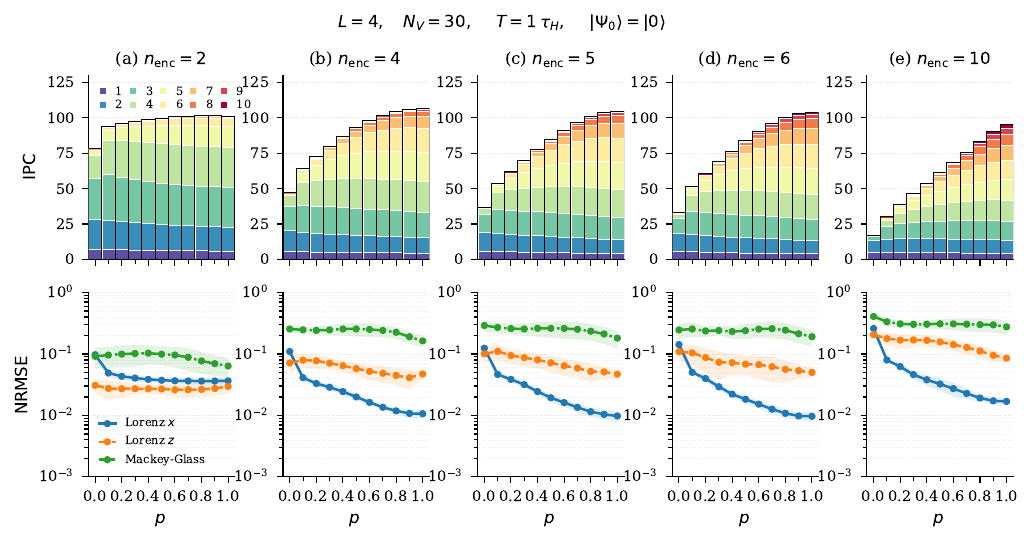}
    \caption{
        Dependence of the total information processing capacity and task
        performance on the hybrid interpolation parameter \(p\) for a reservoir
        of \(L=4\) qubits. Results are shown for
        \(n_{\mathrm{enc}} \in \{2,4,5,6,10\}\).
        The limits \(p=0\) and \(p=1\) correspond to the dissipative and
        data re-uploading protocols, respectively, while intermediate values
        describe the hybrid protocol.
    }
    \label{fig:fig5_L4}
\end{figure}

\begin{figure}[H]
    \centering
    \begin{subfigure}[t]{\textwidth}
        \centering
        \includegraphics[scale = 1]{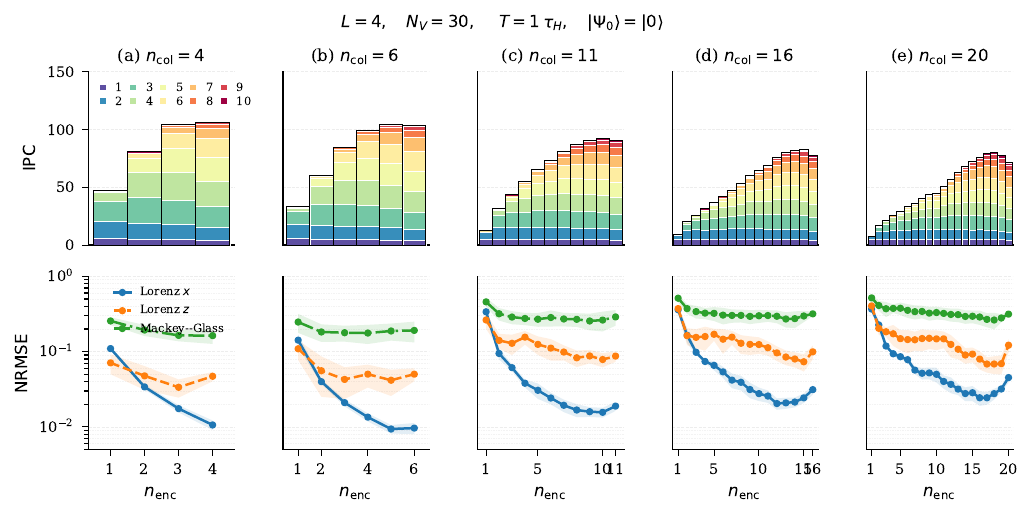}
        %\caption{$T=\tau_H$.}
        \label{fig:hybrid_fixed_L4_T1}
    \end{subfigure}

    \vspace{0.4cm}

    \begin{subfigure}[t]{\textwidth}
        \centering
        \includegraphics[scale = 1]{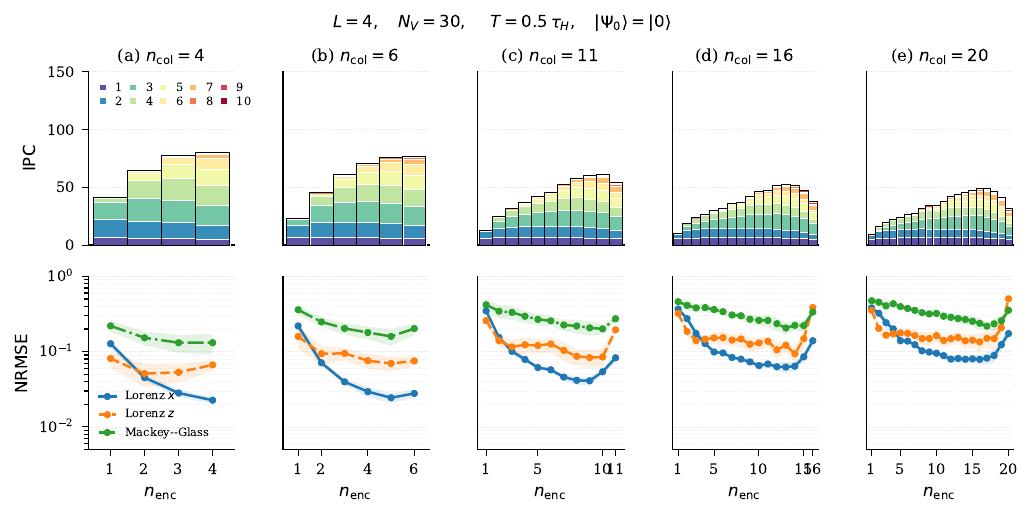}
        %\caption{$T=0.5\tau_H$.}
        \label{fig:hybrid_fixed_L4_T05}
    \end{subfigure}

    \caption{
        Information processing capacity (IPC) and prediction performance of the deterministic hybrid protocol for a four-qubit reservoir ($L=4$) with $N_V=30$. The upper panels show the IPC contributions up to tenth order, while the lower panels show the NRMSE for Lorenz $x$, Lorenz $z$, and Mackey--Glass as a function of $n_{\mathrm{enc}}$ for different numbers of collisions $n_{\mathrm{col}}\in\{4,6,11,16,20\}$. Results are shown for $T=\tau_H$ and  $T=0.5\tau_H$.
    }
    \label{fig:hybrid_fixed_L4}
\end{figure}

In the dissipative protocol, we observe a slight enhancement of the total expressivity for a small number of encodings, in particular around \(n_{\mathrm{enc}}=2\). This raises the question of whether the hybrid protocol can exhibit a regime in which the total expressivity exceeds both the purely dissipative and the full data re-uploading limits. To investigate this, we sweep the interpolation parameter \(p\) for several fixed numbers of encodings, \(n_{\mathrm{enc}} \in \{2,4,5,6,10\}\). The corresponding results are shown in \cref{fig:fig5_L4} for \(L=4\) qubits and in \cref{fig:fig5_L5} for \(L=5\) qubits.

For \(L=4\), we do not observe a pronounced intermediate maximum in the total expressivity as a function of \(p\). For all values of \(n_{\mathrm{enc}}\) considered here, the expressivity increases toward the data re-uploading limit, with the largest values obtained at or close to \(p=1\). A similar tendency is visible in the task performance: the performance on the two Lorenz tasks improves as \(p\) approaches the data re-uploading limit, whereas the Mackey--Glass task remains comparatively insensitive to the interpolation parameter.

The \(L=5\) results show a very similar overall behavior. In this case, however, a small enhancement above the \(p=1\) value is visible for \(n_{\mathrm{enc}}=4\). This feature may originate from contributions of IPC orders beyond those included in our truncation, or it may indicate a genuine but weak intermediate enhancement of the reservoir expressivity. Within the tasks considered here, however, this increase does not lead to a corresponding improvement in task performance.

These results provide an outlook for future studies, in which a hybrid protocol could offer a useful framework combining two distinct mechanisms for modifying reservoir dynamics. The dissipative protocol can enhance already existing IPC contributions, whereas data re-uploading tends to redistribute the information-processing capacity toward higher nonlinear orders. Therefore, an interesting question is whether a protocol can be designed in which data re-uploading first increases the order of the nonlinearities, after which dissipation further enhances the nonlinearities themselves.
For the parameter regimes investigated here, however, the data re-uploading mechanism remains the dominant contribution to the overall enhancement in expressivity. %This indicates that repeated data re-uploading may provide a particularly effective, and still comparatively underexplored, resource for increasing the nonlinear information-processing capabilities of quantum reservoirs.

We note that, for all states considered and for additional multiplexing configurations, we analyzed the behavior of the dissipative, hybrid, and data re-uploading protocols. We did not observe any behavior beyond the trends presented here and those that were expected, such as saturation of task performance for an increasing number of encodings and for larger clock cycles \(T\). This also holds for the more machine-learning-inspired implementation with fixed \(T=\Delta t\, n_{\mathrm{enc}}\). Overall, all results exhibit the same qualitative behavior and can therefore be summarized by the trends discussed here.

 \FloatBarrier
\newpage
\section{Results for four qubits}

%%%%%%%%%%%%%%%%%%%%%%%%%%%%%%%%%%%%%%%%%%%%%%%%%%%%%%%%%%%%
% Bibliography
%%%%%%%%%%%%%%%%%%%%%%%%%%%%%%%%%%%%%%%%%%%%%%%%%%%%%%%%%%%%

%\bibliography{references}